\documentclass[review]{elsarticle}
\usepackage{dsfont}
\usepackage{marvosym}
\usepackage{lineno,hyperref}
\usepackage{amsmath}
\usepackage{tikz}
\usepackage{titlesec}
\usepackage{multirow}
\usepackage{multirow,multicol,makecell}
\usepackage{amssymb}
\usepackage{algorithm}
\usepackage{algorithmic}
\usepackage[utf8]{inputenc}
\usepackage{dsfont}
\usepackage{amsmath}
\usepackage{subfig}
\usepackage{booktabs}
\usepackage{graphicx}
\usepackage{float}
\modulolinenumbers[5]

\journal{Journal of Neurocomputing}

\begin{document}

\begin{frontmatter}

\title{A proactive autoscaling and energy-efficient VM allocation framework using online multi-resource neural network for cloud data center}
\author[a1]{Deepika Saxena \corref{corAuthor}}
\cortext[corAuthor]{Corresponding author}
\ead{13deepikasaxena@gmail.com}
\author[a1]{Ashutosh Kumar Singh}
\ead{ashutosh@nitkkr.ac.in}


\address[a1]{Department of Computer Applications, National Institute of Technology, Kurukshetra, India}
\tnotetext[mytitlenote]{This article has been accepted in Elsevier Neurocomputing Journal © 2020 Elsevier. Personal use of this material is permitted. Permission from
	Elsevier must be obtained for all other uses, in any current or future media, including reprinting/republishing this material for advertising or promotional purposes, creating new collective works, for resale or redistribution to servers or lists, or reuse of any copyrighted component of this work in other works. This work is freely available for survey and citation.}




\begin{abstract}

	   This work proposes an energy-efficient resource provisioning and allocation framework to meet dynamic demands of the future applications. The frequent variations in a cloud user's resource demand leads to the problem of an excess power consumption, resource wastage, performance and Quality-of-Service (QoS) degradation. The proposed framework addresses these challenges by matching the application's predicted resource requirement with resource capacity of VMs precisely and thereby consolidating entire load on the minimum number of energy-efficient physical machines (PMs). The three consecutive contributions of the proposed work are: (1) Online Multi-Resource Feed-forward Neural Network (OM-FNN) to forecast the multiple resource demands concurrently for the future applications, (2) autoscaling of VMs based on the clustering of the predicted resource requirements, (3) allocation of the scaled VMs on the energy-efficient PMs. The integrated approach successively optimizes resource utilization, saves energy and automatically adapts to the changes in future application resource demand. The proposed framework is evaluated by using real workload traces of the benchmark Google Cluster Dataset and compared against different scenarios including energy-efficient VM placement (VMP) with resource prediction only, VMP without resource prediction and autoscaling, and optimal VMP with autoscaling based on actual resource utilization. The observed results demonstrate that the proposed integrated approach achieves near-optimal performance against optimal VMP and outperforms rest of the VMPs in terms of power saving and resource utilization up to 88.5\% and 21.12\% respectively. In addition, OM-FNN predictor shows better accuracy, lesser time and space complexity over a traditional single-input and single-output feed-forward neural network (SISO-FNN) predictor.

\end{abstract}


\end{frontmatter}

\section{Introduction}
 Advances in computing and virtualization technologies have enabled the cost-effective realization of large-scale data centers, which execute large portion of internet applications including business, research and innovations, and social networking etc. Commercial cloud service providers (CSPs) offer elastic computing benefits to the user in the form of variety of computing instances or Virtual Machines (VMs), having different resource capacities at minimum capital investment \cite{buyya2009cloud}. Such a facility allows cloud users to grow and shrink their resource demands and pay accordingly. As the user demand grows, additional VMs are scaled up to satisfy the QoS requirement. Conversely, the VMs are scale down or released, when the resource demand drops down, to save the cost. Further, the rapid changes in the workload demand, results into an inefficient placement of VMs on the PMs and unevenness in the resource distribution at data center. The under-utilized PMs cause resource wastage and excess power consumption \cite{barroso2013datacenter} while over-utilized PMs cause degradation of performance and QoS. Therefore, an intelligent and efficient resource management scheme, is required to reduce the resource wastage, excess consumption of power, and operational cost for the CSPs \cite{kumar2019cloud} while ensuring a satisfactory performance and QoS to the cloud users. 

\par  A VM autoscaling approach enabled with prior estimation of resource utilization and information of an expected variation in the future workload, allows proactive selection of appropriate quantities and types of VMs, to execute the varying applications of the cloud user. In autoscaling of VMs, the applications are proactively assigned to the VMs according to their predicted resource requirement. The types and quantities of the scaled VMs must be sufficient for the execution of user's application with satisfactory performance, yet not  over-estimated to avoid unnecessary cost to the users and resource wastage to the CSP. Hence, a proactive system prepares VMs in anticipation while avoiding latency and performance degradation. Furthermore, effectiveness of the proactive VM autoscaling depends on the potential decisions of the placement of VMs on the PMs. The placement of scaled VMs on minimum and sufficient number of energy-efficient PMs improves performance, minimizes the operational cost by reducing unnecessary power consumption and generates higher revenues by executing more applications. 
\par Significant prior work exists for the cloud resource management, which considers prediction and allocation of the resources exclusively, for example, workload prediction models were presented in \cite{prevost2011prediction}, \cite{kumar2018long}, \cite{kumar2020biphase} etc., resource allocation was discussed in \cite{saxena2016dynamic}, \cite{gao2013multi} etc. However, they work interactively in a real environment where, the workload prediction forecasts the future resource utilization of VMs beforehand, and the resource allocation deals with placement of VMs on the PMs to balance the workload efficiently. In the cloud data center, all these operations work continuously in a cooperative manner to allow an optimized resource allocation. Conversely, the information provided by the resource prediction degrades, if the VMs are not scaled effectively and autoscaling of VMs while ignoring optimization during VM placement, cannot bring the possible power saving and efficient resource utilization. Therefore, a combined 'Proactive VM autoscaling and placement' is a complex and challenging research problem which requires cooperation of all the aforementioned operations to achieve the real benefits of cost and performance optimization for both the cloud user as well as the service provider.

\par 
To address the aforementioned challenges, in this work, the required quantities and types of VMs are proactively determined and adjusted dynamically to meet the future resource demands of the application followed by their placement on energy-efficient PMs. It has been observed that evolutionary neural networks based workload prediction provides improved accuracy over traditional neural networks trained with back propagation algorithm \cite{kumar2020biphase}, \cite{kumar2018workload}. The application of evolutionary algorithms like genetic algorithm and differential evolution for optimization of neural network enhances their learning and prediction capability by allowing extensive exploration and exploitation in multiple directions (or solutions). Therefore, a novel Online Multi-Resource Feed-forward Neural Network (OM-FNN) predictor model based on evolutionary optimization is developed to concurrently predict multiple resource requirement of tasks (sub-unit of application) execution on VMs. The predicted tasks are grouped into clusters as per their resource demands to allow optimal autoscaling of VMs subject to an adequate number and type of VMs. Furthermore, scaled VMs are assigned to energy-efficient PMs by applying proposed multi-objective VM placement approach to achieve an enhanced performance with maximum resource utilization and power saving.  
\subsection{Our contributions} 
The key contributions of the proposed work can be summarized as follows:

\begin{itemize}

	\item A novel \textbf{O}nline \textbf{M}ulti-Resource \textbf{F}eed-forward \textbf{N}eural \textbf{N}etwork ({OM-FNN}) predictor model is proposed to forecast multiple resource utilization concurrently (against traditional feed-forward neural network that works for single resource only) with enhanced accuracy accompanied with Error-Driven Padding (EDP). 
	\item A Tri-adaptive Differential Evolution (TaDE) learning algorithm is developed which explores the search space globally and exploits the population of solutions (networks) to select optimal solution. It is applied for optimization of OM-FNN predictor.
	
	\item An integrated multiple resource usage prediction and clustering of tasks based proactive VM Autoscaling framework is designed to maximize cloud service provider's revenue and bring efficient management of elastic resources in cloud environment.
	
	\item Substantial power saving and improvement in the resource utilization are achieved by exploiting the multiple resource usage prediction, applying successive optimization during task assignment, VM scaling and placement, reducing the number of active servers and VM migrations.
	\item Implementation and evaluation of the proposed framework by using a real benchmark Google Cluster dataset reveals that the proposed work outperforms the state-of-the-art approaches in terms of performance metrics like resource usage prediction, resource utilization and reduction of the power consumption.  

\end{itemize}

\subsection{Organization} The rest of the paper is organized as follows. Section 2 defines the resource management problem. Section 3 provides the recent key contributions categorized into $(i)$ workload prediction approaches, where we have considered the methods based on neural network specifically, $(ii)$ VM Autoscaling based on machine learning approaches and $(iii)$ Multi-objective VM placement approaches based on evolutionary optimization. Section 4 discusses the proposed resource management framework followed by Section 5 and Section 6 which gives the detailed description of proactive VM autoscaling and VM placement respectively. The main algorithm and complexity analysis of the proposed resource management approach is discussed in Section 7. The performance evaluation of the proposed work including the comparison results and analytical remarks are presented in Section 8. Finally, the paper is concluded in Section 9.

\section{Problem definition}
\begin{figure}[!htbp]
	\centering
	\includegraphics[width=0.9\linewidth]{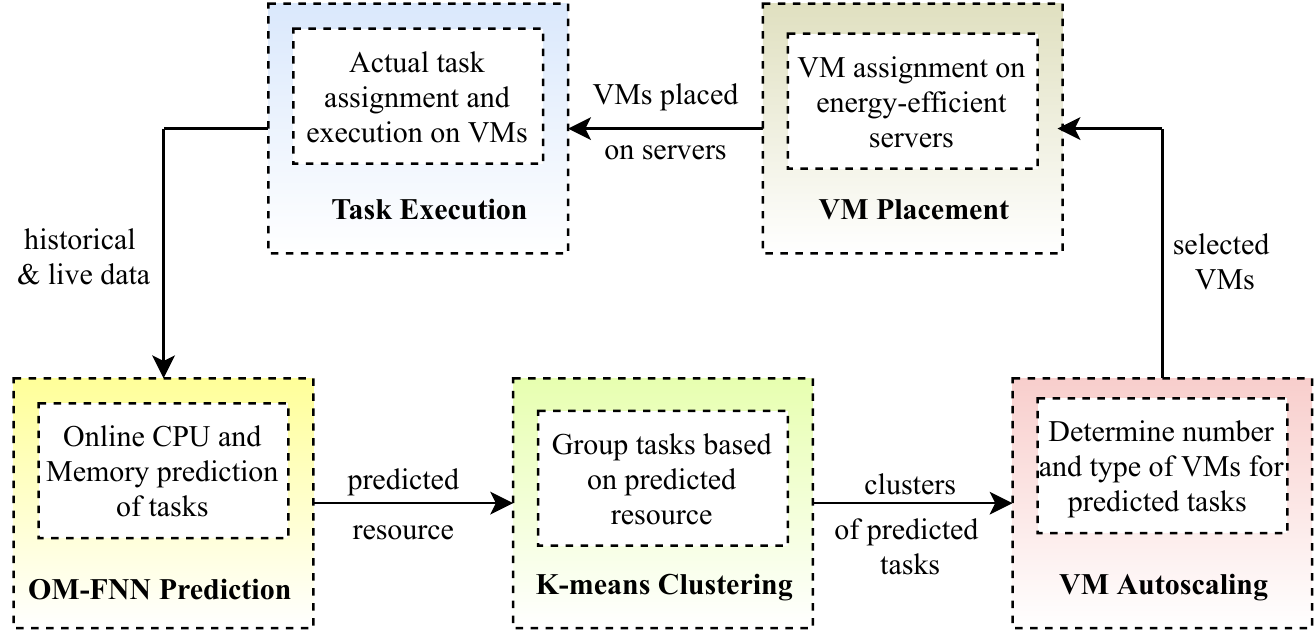}
	\caption{Bird eye view of proposed work }
	\label{fig:flowchart}
\end{figure}

A cloud data center consists of group of servers packed into clusters/racks, where Resource Manager (RM) is responsible for resource distribution, scaling of virtual machines (VMs), jobs and VMs scheduling. It receives request from cloud users and deploys VMs with requested capacity of resources on the appropriate server for the execution of their requests. Let actual resource requirement $\mathds{R}$ of $j_{th}$ VM $v_j$ for different resources \{$r_1$, $r_2$, ..., $r_x$\}, is represented as $\mathds{R}$= \{$\mathds{R}^{v_j}_{r_1}$, $\mathds{R}^{v_j}_{r_2}$, ..., $\mathds{R}^{v_j}_{r_x}$\}. RM utilizes the prior resource requirement information, received from the predictor system for an efficient load balancing to (i) avoid delay during resource distribution upon arrival of the user request, (ii) reduce resource wastage by scaling the adequate number and type of VMs and (iii) allow resource provisioning and deploy VMs on the minimum number of active servers. The predicted information of different resources plays a very crucial role for RM, as it helps in making load management decisions, scaling down the operational cost and enhancing the financial gain for CSP. Let the predicted resource requirement $\hat{\mathds{R}}$ is represented as $\hat{\mathds{R}}^{v_j}$=\{$\hat{\mathds{R}}^{v_j}_{r_1}$, $\hat{\mathds{R}}^{v_j}_{r_2}$, ..., $\hat{\mathds{R}}^{v_j}_{r_x}$\} where, $\hat{\mathds{R}}^{v_j}_{r_1}$ is predicted resource requirement of $j^{th}$ VM $v_j$ for  $i^{th}$ resource $r_i$. The prediction error of different resources of $j^{th}$ VM $v_j$ is denoted as $\xi$=\{$\xi^{v_j}_{r_1}$, $\xi^{v_j}_{r_2}$, ..., $\xi^{v_j}_{r_x}$\}. However, deploying separate predictor system for each resource or using the same predictor system again for the distinct resource prediction raises an unnecessary overhead, time and space complexity. Conceptually, for prediction of $x$ resources, the complexities are raised by $x$ times, which is impractical. Hence, the objectives of the proposed work are (i) to reduce these complexities, by developing and deploying a single multi-resource predictor that can predict the utilization of multiple resources concurrently while minimizing the prediction errors $\xi$=\{$\xi^{v_j}_{r_1}$, $\xi^{v_j}_{r_2}$, ..., $\xi^{v_j}_{r_x}$\} associated to them. (ii) to maximize CSP's revenue by utilizing the predicted information for an adaptive autoscaling of VMs while minimizing resource wastage. (iii) an energy efficient resource distribution during VM placement and avoiding Service-Level Agreement (SLA) violation. Fig. \ref{fig:flowchart} gives a bird eye view of proposed work and highlights our consecutive contributions for efficient resource management at cloud data center.

\section{Recent Key Contributions} 
\subsection{Workload prediction approaches based on Neural Network} 
A future workload prediction technique based on Back-propagation training of three-layered neural network was developed in \cite{prevost2011prediction} that predicted workload with good accuracy on NASA HTTP web log traces for prediction interval upto 60 seconds. However, it was observer that prediction accuracy decreases with increasing prediction interval. Later, an artificial neural network based workload prediction model, trained with self-adaptive differential evolution (SaDE) algorithm was presented in \cite{kumar2018workload}. The application of evolutionary algorithm has outperformed the Backpropagation approach \cite{prevost2011prediction} for neural network optimization, because of its exploration and exploitation capabilities. Kumar et al. had proposed a dynamic resource scaling by using neural network and blackhole learning algorithm based workload prediction approach in \cite{kumar2016dynamic}, which has further outperformed Backpropagataion. A workload prediction approach based on evolutionary neural network was presented in \cite{mason2018predicting}. The approach implemented particle swarm optimization, differential evolution, and covariance matrix adaptation evolutionary strategy learning algorithms and compared their performance. A fine-grained host load predictive models based on long short-term memory model in a recurrent neural network (LSTM-RNN) were presented in \cite{song2018host} and \cite{kumar2018long}. Though the LSTM-RNN model learns long-term dependencies and produce high accuracy for host loads, they suffer from long computation time during training because of the usage of backpropagation algorithm between recurrent layers. To allow high capability of learning and better accuracy in less time, multi-layered neural networks with multi-valued neurons (MLMVN) prediction model was proposed in \cite{qazi2018cloud}. This work applied a complex-valued neural network \cite{yuan2019synchronization}, with derivative-free feed-forward learning algorithm based workload prediction model, that produced better forecasting accuracy than LSTM-RNN approach. Tseng et al. proposed a Genetic Algorithm based workload prediction approach in \cite{tseng2017dynamic}, for resource management which improves the average utilization and energy consumption. 
\par Recently, machine learning approaches based VM prediction models were presented in \cite{moghaddam2020embedding} for energy efficient VM consolidation. The traditional prediction models applied in this work included Linear regression, Multi-layer Perceptron, Support Vector Regression, Decision Tree and Boosted decision tree regression. They applied cross-validation to select appropriate prediction model for each particular VM, which is an extreme time consuming process for real world cloud data center. Moreover, they had predicted CPU utilization only for VM prediction and ignored to predict memory requirement of VMs. Wiener-Filter based prediction approach was applied in \cite{dabbagh2016energy} for resource prediction of a VM and to detect overload occurrence on a server. Separate prediction model was employed to predict each resource utilization at a VM, which means multiple predictor systems were installed at single VM. Recently, an ensemble learning based VM resource request prediction was presented in \cite{kumar2020ensemble}, where the authors have applied Blackhole learning based evolutionary algorithm for training of feed-forward neural network.     
        
\subsection{ Scaling of VMs }
Reinforcement learning based Fuzzy approaches for autoscaling of VMs are presented in \cite{arabnejad2017comparison} to reduce application cost and guaranteed SLA. It comprises of Fuzzy Q-learning (FQL) and Fuzzy SARSA-learning (FSL). FQL is an off-policy approach, in which Q-learning is independent of the policy currently followed. FSL is an on-policy which incorporates the actual agent's behavior and leads to faster learning. Both approaches were capable of handling various load traffic situations, sudden and periodic, and on-demand resource delivery. Moghaddam et al. proposed an anomaly-based cause aware auto-scaling (ACAS) framework for VMs in \cite{moghaddam2019acas}. It utilized isolation-trees for detection of a low overhead anomaly and combined it with a cause identification procedure to allow an appropriate auto-scaling solution, considering the nature of the anomaly. 
\par A VM autoscaling method for online-malware detection was proposed in \cite{abdelsalam2019online}. It employed process-level performance metrics to model a Convolutional Neural Network (CNN) which was trained on samples of VMs during autoscaling to allow detection of malware at run-time. Guo et al. \cite{guo2018online} have presented VM auto-scaling for hosting elastic applications of user which changes over time. In this work, a shadow algorithm is proposed that employed a specifically configured virtual queueing mechanism, to dynamically provide an optimal solution that guides the VM auto-scaling and the VM-to-PM packing .
 A Fast launch Event-driven Auto-Tuning (FEAT) of VMs was presented in \cite{novak2019cloud}. It utilized cloud functions available at cloud providers as interim resources to deal with the delay in launching of VMs and an auto-scaling algorithm was adopted without any requirement of pre-specified thresholds, that made it robust against frequently changing workload. 
 Ruiz et al. \cite{ruiz2017rls} had presented a resource adaptation approach that employed proactive memory-based vertical scaling of VMs. 
  An automatic adaptation of VM's computational capabilities was applied according to its resource usage and performance. Their solution was able to maintain the expected performance, while reducing resource wastage. An online controller design built on top of the Xen hyper visor was presented in \cite{yazdanov2012vertical}. The controller allowed the elastic provisioning of applications by applying a collaboration of resource constraints adaptation of VMs and dynamic plugging of new virtual CPUs. This work allowed to reduce the total CPU time as compared to statically allocated CPU with minimization of SLA violation rate and provided stable response time for high priority VM.
 
\subsection{Multi-objective VM Placement}
Many population based approaches like genetic algorithm (GA), swarm intelligence such as PSO, ACO and Firefly optimization algorithms  have been applied for VM placement \cite{donyagard2019multiobjective}. The GA based approaches are applied in numerous previous work including \cite{xu2010multi}, \cite{wang2013new},  \cite{singh2019secure}, \cite{tseng2017dynamic}, and \cite{liu2014new}. 
Recently, Singh et al. \cite{singh2019secure} presented secure and energy aware load balancing (SEA-LB) framework based on GA approach to introduce the security concept by minimizing number of conflicting servers along with power saving and efficient resource utilization. The drawback is that the role of VM migration was ignored during load balancing. The limitation of GA based VM placement is that it often leads to premature convergence. 
\par Sharma et al. \cite{sharma2016multi} presented an Euclidean distance based multi-objective energy efficient VM placement on servers at cloud data center. The authors proposed HGAPSO algorithm by combining genetic algorithm (GA) and particle swarm optimization (PSO) to minimize resource wastage and SLA violation during VM allocation. GA helps in migration of VMs from source to target server and PSO assists GA in selecting optimal target server by allowing VM placement from non-energy efficient to energy-efficient server. The PSO based approach encodes VM allocations as particle velocity vector where a bit value is 0 if the server is in sleep mode and bit value is 1 for an active server hosting one or more VMs. This method is suitable for homogeneous VM placement because the bit value of velocity vector depends on presence or absence of VMs only and do not perfectly encode for number and type of VMs, hence not suitable for heterogeneous environment. Ant Colony Optimization (ACO) based multi-objective VM consolidation is presented in \cite{liu2014energy}, \cite{gao2013multi}, \cite{ferdaus2014virtual} and \cite{zheng2016virtual}.  Liu et al. used ACO \cite{liu2016energy} for the assignment of VMs on servers from a global optimization perspective by pheromone deposition which guides the artificial ants towards promising solutions and group candidate VMs together. There is a bond among the VMs on the same server and records good VM groups through learning from historical experience. Moreover, Gao et al., \cite{gao2013multi} proposed multi-objective ant colony optimization (ACO) for optimal VM placement and efficient power consumption. The drawback of ACO is that it depends on quantity of pheromones to search optimal solution in search space and it is unsuitable to recursively improve the resource utilization.
\par It has been observed that considerable research work is available that assists in cloud resource management at different levels viz. resource prediction, VM scaling and VM placement. However, in real cloud environment, these operations interactively works at unified platform, which is lacking in previous methods of resource management. In the light of above works, the proposed framework contributes an interactive collaboration of these necessary operations to achieve an efficient and pragmatic management of cloud resources. Best of authors knowledge, the novelty of proposed framework can be realized at various operational levels as: (i) An online and multi-input and multi-output predictor is developed that can predict multiple resources at once. It is trained by applying proposed TaDE optimization algorithm with historical data and retrained with live data periodically to allow accurate prediction online. Further EDP feature is added (as a safety measure to mitigate effect of any prediction error) to predicted output each time. (ii) Utilization of multiple resource prediction of tasks to group them into clusters and accordingly determine the required number and type of VMs to be scaled in future. (iii) Multi-objective evolutionary algorithm is applied for energy-efficient allocation of scaled VMs on available PMs.
  
\section{Proposed resource management framework}

\begin{figure}[!htbp]
 	\centering
 	\includegraphics[width=1.2\linewidth]{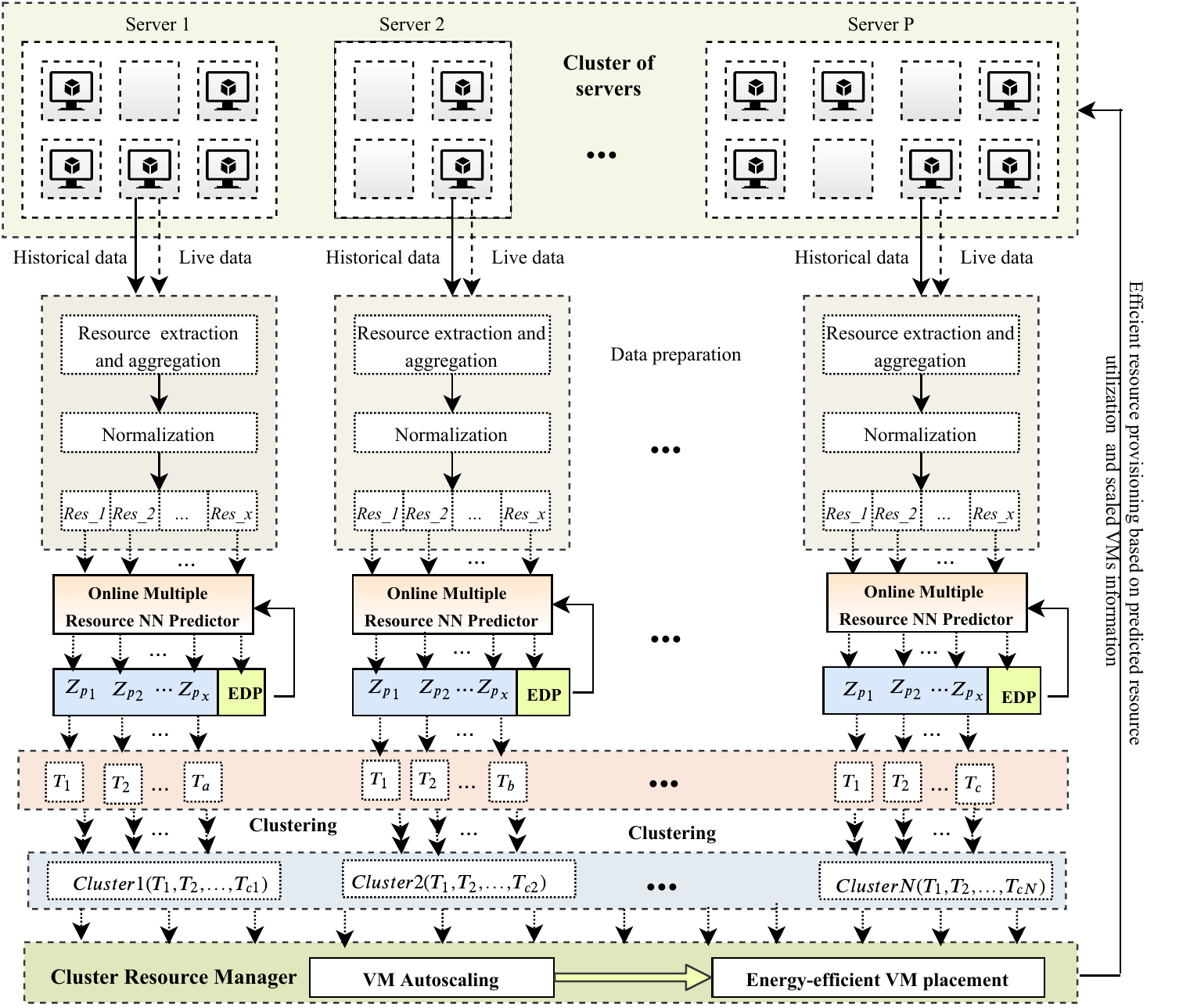}
 	\caption{ Workflow of proposed predictive and autoscaling framework}
 	\label{fig:detailed-workflow-of-proposed-predictive-framework}
 \end{figure}

 Fig. \ref{fig:detailed-workflow-of-proposed-predictive-framework} outlines the detailed workflow of proposed framework. A cluster of servers is shown at the top, where VMs of different users are deployed for execution of their applications. The allocated VMs are shown in occupied blocks and the vacant blocks show released/de-allocated VMs on the server. The resource utilization information of live and historical task execution on a particular VM helps to predict the resource requirement of the future task on the respective VM. Therefore, a distinct multi-resource predictor is dedicated to each VM and an exclusive data preparation occurs before prediction.  

Data preparation is performed in the two steps including attribute extraction and aggregation followed by normalization. From both historical and live data, useful attributes (related to resources viz. CPU and memory utilization of a task at particular VM) are extracted and aggregated per unit time (for eg. 5 min for proposed work) to forecast the resource usage information of a task during next prediction interval. The aggregated values of each attribute/resource are normalized by applying Eq. \ref{eqn:Normalization} 
\begin{equation}
\label{eqn:Normalization}
\hat{D}= \frac{ D_i- D_{min}}{D_{max}-D_{min}}
\end{equation}
where $D_{min}$ and $D_{max}$ are the minimum and maximum values of the input data set, respectively. The normalized vector $\hat{D}$ is a set of all normalized input data values for a particular resource usage. For each resource, separate normalized vector is computed and multiple (depending upon the number of resources considered) input vectors are fed into the input layer of OM-FNN. The previous prediction error score i.e. Error Driven-Padding (EDP) is padded with training input vector to forecast the multiple resources utilization for the next periodic session. The future task resource requirements are generated as predicted output.
\par The clustering operation is applied on the predicted resource requirement information to classify the future tasks (to be executed on VMs) based on their computational and storage complexities. The task classification helps in prior estimation of an adequate number and type (based on configuration) of VMs needed to execute the future workload. Therefore, the collaboration of multiple resource prediction and clustering assists in 'Proactive Autoscaling' of VMs periodically. Further, the chosen scaled VMs are proactively placed on the selected energy-efficient servers which allow maximum resource utilization and power saving. When user application requests are received, the scaled VMs are loaded with tasks (or sub-units of the application) for execution. The entire process allows an efficient resource provisioning which repeats periodically over consecutive prediction intervals. The detailed description of OM-FNN predictor and clustering based VM autoscaling are given in subsequent section \ref{ProactiveVMautoscaling} i.e. proactive VM autoscaling followed by discussion of  energy-efficient VM placement in section \ref{VMplacement}.    

\section{Proactive VM autoscaling}\label{ProactiveVMautoscaling}

The proactive autoscaling of VMs is accomplished in two successive steps including online multiple resource prediction and clustering based VM autoscaling.
 \subsection{ Online Multiple Resource Prediction} 
\begin{figure}[!htbp]
 	\centering
 	\includegraphics[width=1.1\linewidth]{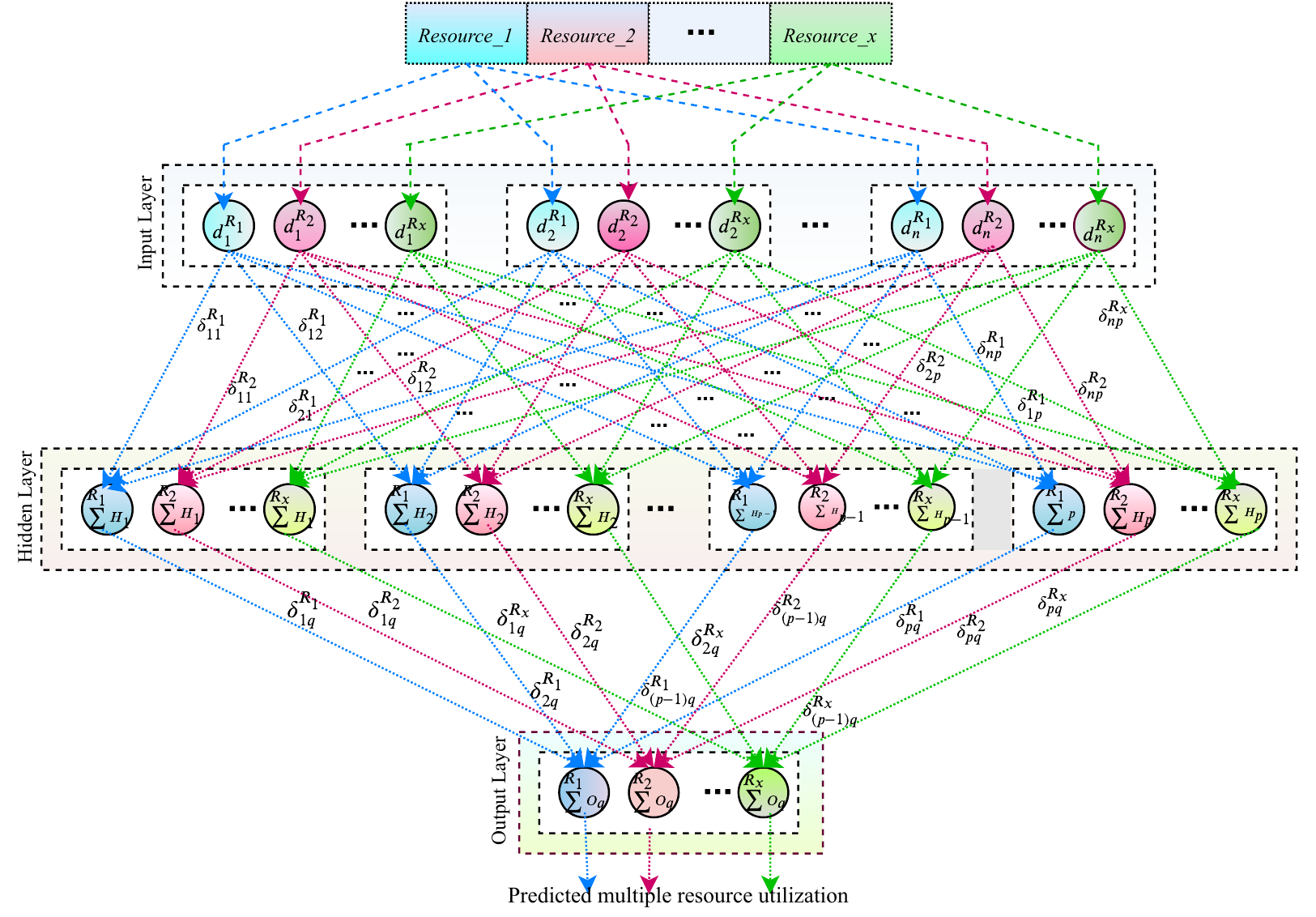}
 	\caption{Online Multi-Resource Feed-forward Neural Network (OM-FNN) predictor  }
 	\label{OM-FNN}
 \end{figure}

 An Online Multi-resource Feed-forward Neural Network (OM-FNN) predictor is developed by modifying functionality of an existing feed-forward evolutionary neural network that receives input and predicts output based on multiple resources (or attributes). It performs joint classification and prediction operations during optimization process and classifies the concurrently predicted information of multiple resources utilization as the output. Instead of conventional nodes, there are sets of nodes at each layer as shown in Fig. \ref{OM-FNN} where, the input, hidden and output layers have $n$, $p$ and $q$ sets of nodes. Let there are $x$ different resources, represented as \{$Resources\_1$, $Resources\_2$, ..., $Resources\_x$\}. The input data vector is \{\{$d_1^{\mathds{R}_1}$, $d_1^{\mathds{R}_2}$, ..., $d_1^{\mathds{R}_x}$\}, \{$d_2^{\mathds{R}_1}$, $d_2^{\mathds{R}_2}$, ..., $d_2^{\mathds{R}_x}$\}, ..., \{$d_n^{\mathds{R}_1}$, $d_n^{\mathds{R}_2}$, ..., $d_n^{\mathds{R}_x}$\}\} where $d_i^{\mathds{R}_j}$ is the input data point given to $j^{th}$ node of $i^{th}$ set which specify $i^{th}$ previous utilization of $j^{th}$ resource. Each network is represented as $\Phi$ and its size can be defined as $L = \sum_{i=1}^{x}{(n + 1) \times p + p \times q}$, where, one bias input is also added with $n$ inputs. The data points of each network vector are generated randomly with an uniform distribution in the range [-1, 1]. The combination of $n-1$ historical and one live resource utilization information are given as input to train and retrain the OM-FNN predictor periodically to forecast the future resource requirement for the next $(n+1)^{th}$ instance. Hence, there are $n$ sets of $x$ nodes each in the input layer. Similarly, there are $p$ and $q$ sets of $x$ nodes at the hidden and output layer represented as \{\{$\sum{H_1}^{\mathds{R}_1}$, $\sum{H_1}^{\mathds{R}_2}$, ..., $\sum{H_1}^{\mathds{R}_x}$\}, \{$\sum{H_2}^{\mathds{R}_1}$, $\sum{H_2}^{\mathds{R}_2}$, ..., $\sum{H_2}^{\mathds{R}_x}$\}, ..., \{$\sum{H_p}^{\mathds{R}_1}$, $\sum{H_p}^{\mathds{R}_2}$, ..., $\sum{H_p}^{\mathds{R}_x}$\}\} and  \{$\sum{O_q}^{\mathds{R}_1}$, $\sum{O_q}^{\mathds{R}_2}$, ..., $\sum{O_q}^{\mathds{R}_x}$\} respectively. The network connections between input and hidden layers are denoted as $\delta_{ij}^{\mathds{R}_k}$ where, $k^{th}$ node of $i^{th}$ set in input layer is connected to $k^{th}$ node of $j^{th}$ set in hidden layer. Likewise, $\delta_{ij}^{\mathds{R}_k}$ represents network connections between the hidden and output layer such as, $k^{th}$ node of $i^{th}$ set in hidden layer is linked to $k^{th}$ node of $j^{th}$ set in the output layer.

 The fitness of each network is evaluated by computing RMSE score ($\xi$) as stated in Eq. \ref{rmse}, where $m$ is a number of data samples, $Z_{a}$ and $Z_{p}$ are actual and predicted output respectively.  Since, prediction accuracy is inversely proportional to RMSE, the purpose is to minimize the fitness function. 
 \begin{gather}\label{rmse}
 \xi = \frac{1}{m}\sum_{i=1}^{m}(Z_{a}-Z_{p})^2 
 \end{gather}

 For fine-grained and accurate anticipation of resource utilization, an OM-FNN predictor is dedicated to each VM. OM-FNN forecast useful information by extracting and correlating the patterns from input data. The workload arrival at cloud data center is dynamic with sudden peaks and falls over the time. To learn this dynamically changing behaviour of cloud workload, a dynamic and adaptive optimization algorithm is needed to train online multi-resource neural network predictor. Though lots of learning algorithms like differential evolution, genetic algorithm are available for training of evolutionary neural networks. However, to improve the learning/optimization and adaptation capabilities of differential evolution algorithm, a tri-adaptive differential evolution (TaDE) algorithm is developed that allows training of OM-FNN predictor with improved accuracy and faster convergence (but avoids pre-mature convergence). 
  The consecutive steps of TaDE algorithm are as follows:
 \begin{itemize}
 	\item[1.] It begins with initialization of $N$ number of networks, number of maximum generations ($Gmax$), mutation rate and crossover rate.
  	\item[2.] The fitness of each network is evaluated on training data by applying an error estimation function i.e. Root Mean Square Error (RMSE). 
 	\item[3.] For each generation, mutation selection probability $msp$ is generated, to select one of the three optional mutation schemes and apply on each network, to generate its mutant vector. 
 	\item[4.] Mutation is followed by crossover, in which crossover selection probability $csp$ selects one of the two crossover schemes to generate new offspring for the next generation.
 	\item[5.] Fitness of each offspring vector is evaluated by applying RMSE and select the optimal solution to proceed in next generation.
 	\item[6.]  The control parameters viz. crossover and mutation rates are adaptively tuned during evolutionary optimization.
 \end{itemize}
Overall, the adaptation is applied in three dimensions including, mutation, crossover and control parameters. 
 \begin{itemize}
 	\item 

\textit{Mutation adaptation} Three mutation strategies opted for proposed work are $DE/best/1\quad (MS_1) $, $ DE/current-to-best/1 \quad (MS_2) $ and  $DE/rand/1 \quad (MS_3) $. The mutation strategies $MS_1$, $MS_2$ shown in Eq. \ref{eqn:mutation1} and \ref{eqn:mutation2} tend to be greedy as they exploit the best individual to generate mutant vectors while $MS_3$ stated in Eq. \ref{eqn:mutation3} is applicable for raising population diversity.

 \begin{equation} 
 \Lambda_i^j = \Phi_{best}^j + \mu_i \times (\Phi_{r1}^j -\Phi_{r2}^j)\label{eqn:mutation1}
 \end{equation}
 
 \begin{equation}
 \Lambda_i^j = \Phi_i^j +  \mu_i \times (\Phi_{best}^j -\Phi_i^j) +  \mu_i \times (\Phi_{r1}^j -\Phi_{r2}^j)\label{eqn:mutation2}
 \end{equation}
 \begin{equation}
 \Lambda_i^j = \Phi_{r3}^j + \mu_i \times (\Phi_{r1}^j -\Phi_{r2}^j) \label{eqn:mutation3}
 \end{equation}
 where $\Lambda_i^j$ and $\Phi_i^j$ depicts $i^{th}$ mutant and current vector solution of $j^{th}$ iteration, respectively. The term $\Phi_{best}^j$ is the best solution found so far, till $j^{th}$ generation and $r1$, $r2$ and $r3$ are mutually distinct random numbers in the range [1, N]. To decide the mutation scheme for current iteration, a random probability vector $msp$ is generated.
 The mutation strategy selection is represented as $\wp_m $ in Eq. \ref{mutation_probability}, as follows: 
 \begin{equation}\label{mutation_probability} 
 \wp_m =
 \begin{cases}
 MS_1, & {If(0< msp_i \leq \Gamma_1 )} \\
 MS_2, & {If(\Gamma_1  < msp_i  \leq \Gamma_1 + \Gamma_2 )} \\
 MS_3, & {\text{otherwise}}  
 \end{cases}
 \end{equation}
 where $\Gamma_1$, $\Gamma_2$ and $\Gamma_3$ are the probabilities for opting the  $MS_1 $,  $MS_2 $ and $MS_3$ mutation techniques respectively. In reported experiments, initially $\Gamma_1$=$\Gamma_2$= 0.33, $\Gamma_3$=0.34, so that each mutation scheme get equal chance of selection. 
\item  \textit{Crossover adaptation } After the mutation process, crossover is applied to mutant vector $\Lambda_i^j$, and its corresponding current target vector $\Phi_i^j$, in order to produce new solutions called as offspring $\chi_i^j$ that is $i^{th}$ solution of $j^{th}$ generation. 
 \begin{itemize}

   \item {\textit{Uniform crossover}}\\
   In this approach, crossover occurs at gene level instead of segment level, a random number $\Re$ in the range [0, 1] is generated for each gene of parent chromosome. If the crossover rate $CR_{i}^j$ for $j_{th}$ generation and $i_{th}$ solution (parent) vector is more than the random value generated for the gene, then the gene values of the two parents will be exchanged, otherwise same gene continue in production of  new offspring. This crossover technique allows exploration of both parent chromosome at fine-grained level to produce two new offspring. In each of the above crossover method, two child chromosomes are produced at each generation. We evaluate the fitness of both the children and then select the child having maximum fitness value (least error score) to proceed in the successive generation \cite{pavai2017survey}. Eq.\ref{uniform} shows uniform crossover operation.
   \begin{gather}\label{uniform}
   \chi_i^j=\begin{cases}
   \Lambda_i^j  & {If(\Re\in(0,1) \leq CR_{vi}^j )} \\
   \Phi_i^j   & {\text{otherwise.}} 
   \end{cases}
   \end{gather}

 	\item {\textit{Heuristic crossover}}\\
 	Heuristic crossover compares fitness value of both parent chromosomes and finds the parent with better fitness value to produce a new offspring as depicted in Eq. \ref{heuristic}. It can bring significant diversity in search space, that adds promising genetic material by generating new offspring more closer to parent with better fitness value \cite{wright1991genetic}. Therefore, this crossover operator improves the exploitation capability of DE learning algorithm. 
 	Let two parent vectors are selected, and parent having better fitness and other parent vector, are denoted as ($\Lambda_{better}$) and ($\Lambda_i$) respectively. The offspring $\chi_i^j $ is generated by applying Eq. \ref{heuristic}   
 	\begin{gather}\label{heuristic}
 	\chi_i^j = CR_{vi}^j*(\Lambda_{better}-\Lambda_{i}) + \Lambda_{better}  
 	\end{gather}
 
  \end{itemize}

 Let $\Omega_1$ and $\Omega_2$ be the probabilities for selecting the uniform and heuristic crossover strategies respectively. Similar to  selection of mutation scheme, roulette wheel selection policy, given in Eq.\ref{crossover_probability} is applied to select appropriate crossover strategy. The terms $\wp_c$ represents crossover selection, $csp$ is crossover selection probability.
\begin{equation}\label{crossover_probability}
\wp_c =
\begin{cases}
Uniform crossover, & {If(0< csp_i \leq \Omega_1 )} \\
Heuristic crossover, & {\text{otherwise}}  
\end{cases}
\end{equation}
 
\item  \textit{Control parameters adaptation }
 The convergence speed of adaptive DE depends on the tuning of control parameters crossover rate ($CR$) and mutation rate ($MR$). Before the next generation evolution begins, we count the number of candidates updated in previous generation i.e. $g$. The control parameters $CR$ and $MR$ are updated according to Eq. \ref{mutation control} and \ref{crossover control}, where, $\theta_m$ and $\theta_c$ are uniform random numbers in the range [0, 1], $MR_i^{j+1}$ is mutation rate for $i^{th}$ population vector in the next generation, $MR_l = 0.1$ and $MR_u = 0.8$ are lower and upper bounds for mutation respectively. Similarly, $CR_i^{j+1}$ is the next generation crossover rate of $i^{th}$ population vector, $CR_l = 0.1$ and $CR_u = 0.5$ are the lower and upper limits for crossover operator \cite{wang2010multi}. If the value of $g$ is less than $Z$, we set $Z$ equals to 2:5 of original population size, which means if atleast two-fifth members of the total population is not updated with last values of $MR$ and $CR$, then mutation and crossover rates are regenerated. This will prevent condition of premature convergence. The number of generations elapsed in upgrading values of $MR$ and $CR$ is known as "learning period" for control parameters.
 
 \begin{gather}\label{mutation control}
 MR_i^{j+1}=\begin{cases}
 MR_l + \theta_m(MR_u - MR_l) & {(g\leq Z)} \\
 MR_i^j  & {(otherwise.)} 
 \end{cases}
 \end{gather}
 
 \begin{gather}\label{crossover control}
 CR_i^{j+1}=\begin{cases}
 CR_l + \theta_c(CR_u - CR_l) & {(g\leq Z)} \\
 CR_i^j  & {(otherwise.)} 
 \end{cases}
 \end{gather}
 
 During each generation or epoch, following mutation and crossover, we keep track of the number of candidates successfully reaching the next generation denoted as $sm_1$, $sm_2$, and $sm_3$ for three different mutation strategies. Similarly, $fm_1$, $fm_2$, and $fm_3 $  records the number of candidates failed to reach the next generation.
 The probabilities of successful offspring generated by the $DE/random/1$, $ DE/best/1 $ and $DE/current-to-best/1$ mutation techniques are computed as  $ \rho_1 $,  $ \rho_2 $, and  $ \rho_3 $ shown in Eq. \ref{success-failure}.
 
 \begin{gather}
 \begin{gathered}\label{success-failure}
 b=2(sm_2sm_3 + sm_1sm_3 + sm_2sm_3) + fm_1(sm_2 + sm_3)\\ + fm_2(sm_1 + sm_3) + fm_3(sm_1 + sm_2) \\
 \rho_1= \frac{sm_1(sm_2 + fm_2 + sm_3 + fm_3 )}{b}  \\
 \rho_2= \frac{sm_2(sm_1 + fm_1 + sm_3 + fm_3 )}{b}  \\
 \rho_3= 1- (\rho_1 +\rho_2 )
 \end{gathered}
 \end{gather}
 The probabilities of successful offspring generated by the heuristic and uniform crossover strategies are computed as $\sigma_1$ and  $ \sigma_2 $ shown in Eq. \ref{cs-selection}. Similar to mutation, for crossover too, we keep track of the number of successful and failure candidates denoted as $cs_1$, $cs_2$ and $cf_1$, $cf_2$ respectively reaching the next generation helps to compute $ \sigma_1 $ and  $ \sigma_2 $.
 
 \begin{gather}
 \begin{gathered}\label{cs-selection}
 c=2(cs_2+ cs_1)+ cf_1 \times cs_2 + cf_2 \times cs_1 \\
 \sigma_1= \frac{cs_1(cs_2 + cf_2)}{c}  \\
 \sigma_2= 1- \sigma_1
 \end{gathered}
 \end{gather}

  \end{itemize}
 Finally, successful candidates are selected on the basis of fitness function by applying Eq. \ref{rmse}. The population for the next generation is selected using greedy approach in the form of survival of fittest concept using Eq. \ref{fitness}, where $\delta_i^{j+1}$ is selected candidate for next generation, $\chi_i^j $ is the solution generated after crossover and $\delta_i^j$ is a current candidate solution. The operational summary of TaDE learning algorithm for the online multiple resource forecast system is given in Algorithm \ref{algo-prediction}.
 
 \begin{gather}\label{fitness}
 \delta_i^{j+1}=\begin{cases}
 \chi_i^j & {(fitness(\delta_i^j) \leq (fitness(\nu_i^j))} \\
 \delta_i^j   & {(otherwise.)} 
 \end{cases}
 \end{gather}
 \begin{algorithm}[!htbp]
 	\caption{Proposed TaDE learning algorithm for the online multiple resource forecast system ( )}
 	\label{algo-prediction}
 	\begin{algorithmic}[1]
 		\STATE Initialize crossover and mutation rates, $\Gamma_1$=$\Gamma_2$=0.33, $\Gamma_3$=0.34, $Gmax$ 		
 		\STATE Initialize $N$ networks of size $L$ randomly such as $L =(p+1)\times q + (q\times r) = q(p+r+1) \Leftarrow q(p+2)$		
 		\STATE Evaluate each network on training data using fitness function Eq. \ref{rmse}
 		\FOR{each generation $j^{th} \in G_{max}$} 
 		\STATE Generate vector $msp$ for $N$ networks $\in$ [0,1]
 		\FOR{ each $i^{th}$ network} 
 		\STATE Generate $r_1 \ne r_2 \ne r_3 \ne i \in [1,N] and K_{rand} \in [1,L]$
 		\IF {$0 \textless msp_i \leq  \Gamma_1$}
 		\STATE  Apply$ DE/rand/1$
 		\ELSIF {$\Gamma_1 \textless msp_i \leq (\Gamma_1 + \Gamma_2)$}
 		\STATE  Apply $DE/best/1$ 
 		\ELSE
 		\STATE  Apply $DE/current-to-best/1$ 
 		\ENDIF
 		\STATE Follow steps 7-14 to select and apply either uniform or heuristic crossover	
 		\ENDFOR
 		\STATE Evaluate updated network using error estimation function i.e. Eq. \ref{rmse}
 		\STATE Select participants for next generation using Eq. \ref{fitness}
 		\STATE Update $\Gamma_1$, $\Gamma_2$, $\Gamma_3$ after fixed number of generations
 		\STATE Regenerate mutation and crossover rates by applying Eqs. \ref{mutation control} and \ref{crossover control} respectively.
 		\ENDFOR
 		
 	\end{algorithmic}
 \end{algorithm}

 \textit{Termination condition:} The termination criteria for iterative optimization by applying TaDE algorithm is either of the two: Either the number of iterations becomes greater than maximum iterations ($Gmax$), or when their is no improvement in successive iterations signifying that convergence point has reached. 
  \subsubsection{Error-driven padding (EDP):} Although the proposed on-line predictor is capable of anticipating resource demands with closer precision, still 100\% accuracy cannot be ensured for highly dynamic resource demands. These errors may cause over/under-load and SLA violations. In order to overcome the occurrence of these issues, error-driven precaution margins are padded with predicted resource demand. At $t_{th}$ instance, EDP is computed as $EDP_t=(1-\alpha)\times \xi_{t-1} + \alpha \times  \xi_t $ where $0.5< \alpha \leq 1$. Furthermore, it is to be noted that during EDP computation, more weightage is given to most recent error to improve the accuracy of prediction. Therefore, the improved predicted output becomes $Z_{p} +EDP_t$. 
\subsubsection{An Illustration}
Consider an OM-FNN with 4 input, 3 hidden and 1 output sets of two nodes each having one node for CPU and other for memory to illustrate one iteration of training process. Total number of interconnections or size of network can be computed as $(4 + 1) \times 3 + 3 \times 1 = 18$. Assume four such vectors (each representing one network) from initial population ($\Phi^1$ =\{$\Phi^1_1$, $\Phi^1_2$, $\Phi^1_3$, $\Phi^1_4$\}) are given in Table \ref{initialpopulation}. The initial fitness values of these vectors are computed by applying Eq. \ref{rmse} as shown in Table \ref{fitness1}. 
\begin{table}[!htbp]
	\centering
	
	\caption[Table caption text] {Initial Population ($\Phi^1$)}  
	\label{initialpopulation}
		\resizebox{12cm}{!}{
	\begin{tabular}{cccccccccccccccccc}
		
		\hline
		-0.94 & -0.66& -0.84 & -0.22& -0.126&-0.99&-0.13&-0.15&-0.71&0.06&-0.03&-0.60&0.20&-0.07 & -0.94&-0.42&0.33&0.42\\
		-0.40 & -0.02& 0.56 & -0.97& -0.40&-0.99&0.17&0.26&0.59&0.61&-0.99&-0.29&-0.85 & -0.31&-0.05&0.23&-0.48&-0.36\\
			-0.49 & -0.41& -0.58 & -0.70& -0.59&0.17&-0.94&-0.64&-0.08&-0.02&-0.88&0.18&0.09 & 0.23&0.85&0.32&-0.36&-0.69\\
				-0.72 & -0.89& -0.95& 0.23&0.03&0.11&-0.96&-0.04&0.33&-0.49&-0.86&-0.12&0.17 & 0.17&-0.45&-0.16&0.14&-0.30\\
		\hline
	\end{tabular}}
\end{table}

\begin{table}[!htbp]
	\centering
	
	\caption[Table caption text] {Fitness value of initial population}  
	\label{fitness1}
	\resizebox{5cm}{!}{
		\begin{tabular}{ccccc}			
			\hline
			Fitness value & $\Phi^1_1$ & $\Phi^1_2$ & $\Phi^1_3$ & $\Phi^1_4$\\
			\hline
			$\xi_{CPU}$ & 0.030 &0.023 &0.072&0.002\\
			$\xi_{Mem}$ & 0.027 &0.021 &0.061&0.006\\
				\hline
	\end{tabular}}
\end{table}
The initial values of $msp$, $csp$, $MR$, $CR$ are randomly generated for each member of the population as stated in Table \ref{initialvalues}. The fitness evaluation is followed by the consecutive mutation and crossover operations which generates an offspring population ($\chi^1$) as given in Table \ref{offspringpopulation}. The selection operator (Eq. \ref{fitness}) is applied to produce successive population for the next iteration. 
\begin{table}[!htbp]
	\centering
	
	\caption[Table caption text] {Initial values of following parameters }  
	\label{initialvalues}
	\resizebox{5cm}{!}{
		\begin{tabular}{ccccc}
			
			\hline
			Parameters & $\Phi^1_1$ & $\Phi^1_2$ & $\Phi^1_3$ & $\Phi^1_4$\\
			\hline
			$msp$ & 0.881 &0.846 &0.223&0.754\\
			$csp$ & 0.565 &0.476 &0.823&0.669\\
			$MR$ & 0.002 &0.132 &0.069&0.125\\
			$CR$ & 0.420 &0.732 &0.259&0.203\\
			\hline
	\end{tabular}}
\end{table}
\begin{table}[!htbp]
	\centering
	
	\caption[Table caption text] {Offspring vectors after mutation and crossover ($\chi^1$)}  
	\label{offspringpopulation}
	\resizebox{12cm}{!}{
		\begin{tabular}{cccccccccccccccccc}
			
			\hline
			-0.94 & -0.83& -0.49& -0.22&-0.12&-0.45&-0.13&-0.15&0.71&-0.06&-0.03&0.85&-0.29&-0.57 & -0.14&-0.42&0.33&0.11\\
			-0.93 & -0.92& 0.22 & -0.97& -0.40&-0.99&0.17&0.26&0.59&0.61&-0.99&-0.47&-0.85 & -0.30&-0.40&0.23&-0.48&-0.64\\
			-0.53 & -0.37& -0.88 & -0.70& -0.59&0.48&-0.94&-0.64&-0.08&-0.02&-0.88&0.16&0.47 & 0.05&0.40&0.32&-0.36&-0.41\\
			-0.05 & -0.26& -0.01& 0.23&0.03&0.11&-0.96&-0.04&0.33&-0.49&-0.86&-0.18&0.17 & -0.02&-0.02&-0.16&0.14&-0.11\\
			\hline
	\end{tabular}}
\end{table}
Table \ref{fitness2} evaluates fitness of each member of offspring population and successful candidates reaching second (next) iteration are shown in Table \ref{finalpopulation}. 
\begin{table}[!htbp]
	\centering
	
	\caption[Table caption text] {Fitness value of offspring population}  
	\label{fitness2}
	\resizebox{5cm}{!}{
		\begin{tabular}{ccccc}
			
			\hline
			Fitness value & $\chi^1_1$ & $\chi^1_2$ & $\chi^1_3$ & $\chi^1_4$\\
			\hline
			$\xi_{CPU}$ & 0.022 &0.036 &0.125&0.008\\
			$\xi_{Mem}$ & 0.020 &0.032 &0.109&0.009\\
			\hline
	\end{tabular}}
\end{table}
\begin{table}[!htbp]
	\centering
	
	\caption[Table caption text] {Population for second iteration  ($\Phi^2$)}  
	\label{finalpopulation}
	\resizebox{12cm}{!}{
		\begin{tabular}{cccccccccccccccccc}
			
			\hline
		    -0.94 & -0.83& -0.49& -0.22&-0.12&-0.45&-0.13&-0.15&0.71&-0.06&-0.03&0.85&-0.29&-0.57 & -0.14&-0.42&0.33&0.11\\
			-0.40 & -0.02& 0.56 & -0.97& -0.40&-0.99&0.17&0.26&0.59&0.61&-0.99&-0.29&-0.85 & -0.31&-0.05&0.23&-0.48&-0.36\\
			-0.49 & -0.41& -0.58 & -0.70& -0.59&0.17&-0.94&-0.64&-0.08&-0.02&-0.88&0.183&0.09 & 0.23&0.85&0.32&-0.36&-0.69\\
			-0.72 & -0.89& -0.95& 0.23&0.03&0.116&-0.96&-0.04&0.33&-0.49&-0.86&-0.12&0.17 & 0.17&-0.45&-0.16&0.14&-0.30\\
			\hline
	\end{tabular}}
\end{table}

 \subsection{Autoscaling of VMs}
 The clustering of future tasks is done by applying K-means clustering algorithm, on the basis of predicted data points (i.e. resource utilization of tasks on different VMs). The effective number of clusters (i.e. value of $K$) are decided by applying an Elbow method. K-means partitions the resource utilization of all tasks into $K$ pre-defined distinct non-overlapping clusters or subgroups such that resource utilization of each task belongs to one group only. It iterates to make the inter-cluster resource usage similar, while keeping the clusters as different (far) as possible. The resource utilization of task is assigned to a cluster such that the sum of the squared distance between them and centroid of the cluster is minimum by applying Eq. \ref{eq.cluster}, where $n$ is number of task's resource utilization (data points), $w_{ik}$ defines mapping of $i^{th}$ task predicted resource utilization ($z_i$) in $k^{th}$ cluster and $\mu_k$ is centroid of $k^{th}$ cluster. 
 \begin{equation}
 \label{eq.cluster}
 	G=\sum_{j=1}^{n}\sum_{k=1}^{K}{w_{ik}{|z_i- \mu_k|}^2}
 \end{equation}
 \textit{VM Autoscaling:} The exact number and type of VMs required to execute future workload, is determined by mapping each cluster to an appropriate VM type or size by applying Eq. \ref{eq.autoscale} 
 \begin{equation}\label{eq.autoscale}
 VM^{type}_{selected} =
 \begin{cases}
 v_{small}, & {(z^\mathds{R_{MAX}}_{i} \leq v_{small}^{\mathds{R}} )} \\
 v_{medium} , & {( v_{small}^{\mathds{R}} < z^\mathds{R_{MIN}}_{i} AND z^\mathds{R_{MAX}}_{i} \leq v_{medium}^{\mathds{R}})} \\
 v_{large} , & {(v_{medium}^{\mathds{R}} <z^\mathds{R_{MIN}}_{i}  AND z^\mathds{R_{MAX}}_{i} \leq v_{large}^{\mathds{R}})} \\
 v_{Xlarge} , & {(otherwise.)} 
 \end{cases} \quad \mathds{R} \in \{CPU, memory\}
 \end{equation}
 where $ v_{small}^{\mathds{R}}$, $ v_{medium}^{\mathds{R}}$, $v_{large}^{\mathds{R}}$ and $v_{Xlarge}^{\mathds{R}}$ represents small, medium, large and extra-large types of VM respectively, having capacity of resources $\mathds{R} \in \{CPU, memory\}$ depending on their particular type, and $z^\mathds{R_{MAX}}_{i}$ and $z^\mathds{R_{MIN}}_{i}$ represents maximum and minimum resource utilization of $i^{th}$ cluster. If the maximum resource requirement of a task from $i_{th}$ cluster is lesser or equals to the resource capacity of $v_{small}$, then small type of VM is assigned to the cluster. Likewise, if $( v_{small}^{\mathds{R}} < z^\mathds{R_{MIN}}_{i}  and  z^\mathds{R_{MAX}}_{i}  \leq v_{medium}^{\mathds{R}})$, then $v_{medium} $ is selected for the tasks execution and the required number of VMs is equal to the number of tasks in the respective cluster. 
 
 \section { VM placement} \label{VMplacement}
 In the proposed framework, VMs are placed with respect to maximum resource utilization and minimum power consumption, that can be mathematically stated as Eq. \ref{model}.
 \begin{equation}
 \sum_{i=1}^{p}{S_i} = min\sum_{i=1}^{p}{S_i^{PW}} + max\sum_{i=1}^{p}{S_i^{RU}} \label{model}
 \end{equation}
 where $S_i^{PW}$ and $S_i^{RU}$ are power consumption and resource utilization of $i^{th}$ server respectively. Each VM allocation is feasible only if it satisfies the following constraint given in Eq. \ref{1} where $\omega_{ji}$ shows mapping of $j^{th}$ VM ($v_j$) on $i^{th}$ server ($S_i$). It states that resource requirement of VM ($v_j^{r}$) must be lesser than available resource capacity of server ($S_i^{r}$). 
 \begin{equation}
 \sum_{j=1}^{q}{v_j^{\mathds{R}}} \times \omega_{ji} \leq S_i^{\mathds{R}} \quad   \mathds{R} \in {CPU, Mem}\label{1}
 \end{equation}
 The objective models utilized for optimal VM placement are given in subsequent subsections: 
 \subsection{Resource utilization} Assume $S_i^C$ and $S_i^{M}$ are CPU and memory capacity of $i^{th}$ server. If $i_{th}$ server $S_i$ is active then $\gamma_i=1$, means one or more VMs are placed on it, otherwise, $0$. If server $S_i$ hosts $v_j$, then $\omega_{ji}=1$ otherwise it is 0. For VM $v_j$, CPU and memory utilization are represented as $v_j^{C}$ and $v_j^{M}$ respectively. The resource utilization of data center can be obtained by using Eq. \ref{ru}. Though in formulation, only CPU and memory are considered, it is extendable to any number of resources.
 \begin{equation}
 RU_{dc}= \int\limits_{\substack{t_1\\\mathcal{}}}^{t_2} (\frac{	RU_{dc}^{C} +  RU_{dc}^{M} }{|N|\times \sum_{i=1}^{p}{\gamma_i}})\label{ru}
 \end{equation}
 \begin{equation}
 RU_{dc}^{\mathds{R}}=\sum_{i=1}^{p}{\frac{\sum_{j=1}^{q}{\omega_{ji} \times v_j^{\mathds{R}}}}{S_i^{\mathds{R}}}} \quad \mathds{R} \in {CPU, Mem etc.}
 \end{equation}
       
 \subsection{Power consumption}
 In idle state (sleep mode) CPU works in least frequency mode with reduced clock cycle where power consumption is minimum. On the other hand, power consumption depends on the application assigned for processing and CPU utilization rate $R^{C}$ during busy state.
 Therefore, power consumption for $i^{th}$ server can be formulated as $PW_i$ and total power consumption $PW_{dc}$ during time-interval [$t_1$, $t_2$] is shown in Eq. \ref{power2}.
 \begin{equation}
 PW_{dc} = 
 \int\limits_{\substack{t_1\\\mathcal{}}}^{t_2} (\sum_{i=1}^{p} {([{PW_i}^{max} - {PW_i}^{min}] \times R^{C} + {PW_i}^{idle})})
 \label{power2}
 \end{equation}	
 where ${PW_i}^{max}$, ${PW_i}^{min}$ and ${PW_i}^{idle}$ are maximum, minimum and idle state power consumption for $i^{th}$ server. 
 \subsection{Optimized VM-Allocation approach }The proposed multi-objective VM allocation approach consists of four consecutive stages namely initialization, fitness evaluation, crossover followed by mutation and selection. The VM allocations are represented as chromosomes and the step-by-step procedure is given in Algorithm \ref{algo-mob-lb}. Firstly, $X$ random VM allocations are initialized as $\Psi_i$ (step 1) which represents $i^{th}$ VM placement, subject to $(i\leq X)$ encoded into chromosomes. To evaluate fitness of each chromosome, cost function $\eta(\Psi_g)$ is computed which returns cost values $f_{\Psi_i}^{RU}$ and $f_{\Psi_i}^{PW}$, associated to resource utilization and power consumption respectively which can be evaluated by computing Eq. \ref{ru} and \ref{power2} respectively (step 3). Then, non-dominated sorting is applied by calling Pareto-optimal module, i.e. Algorithm \ref{pareto} to sort each chromosome (i.e. VM allocation) with respect to its dominance level (Algorithm \ref{pareto}: steps 3-13) and put all the non-dominated solutions into pareto-front (Algorithm \ref{pareto}: steps 15-26). If the cost values of chromosome $\Psi_i$ is better with respect to atleast one objective and same or better for rest of the objectives, then $\Psi_i$ dominates chromosome $\Psi_j$. The chromosome with best fitness value is represented as $\Psi_{NDS}$ (Algorithm \ref{algo-mob-lb}: step 4). Further, one-point crossover ($Cr$) and mutation ($\mu$) operations are applied to generate new offspring in order to explore the entire search space for better solution by migrating VMs from non-optimal to selected optimal server where, $cp$ is position of one-point crossover (Algorithm \ref{algo-mob-lb}: step 6). The resultant solutions may be infeasible with respect to VM allocation constraints, which are turned into feasible solutions by re-arranging them. Again, the fitness of updated solutions are evaluated (Algorithm \ref{algo-mob-lb}: step 13,14) and optimal solution is selected by applying multi-objective rank based non-dominating sorting to replace the least fit solution by better ones. Finally, VMs are optimally placed on servers subject to maximum resource utilization with reduced power consumption.

 \begin{algorithm}[htbp]
 	\caption{Multi-objective VM placement algorithm ():}
 	\label{algo-mob-lb}
 	\begin{algorithmic}[1]
 		\STATE Initialize $n$ random VM allocations ($\Psi_1, \Psi_2,...,\Psi_X $). 
 		\FOR {$g={1,2,...,Gmax}$}
 		\STATE  $[f_{\Psi_i}^{RU},f_{\Psi_i}^{PW}] = \eta(\Psi^g)$
 		\STATE $[\Psi_{NDS}= Pareto-optimal(\Psi^g)]$, $\Psi_{best} \leftarrow \Psi_{NDS}[0]$		
 		\FOR {each i=(1,2,...,n) }
 		\STATE $rn= random(1,n)$, $cp= random(1,P)$, where $cp$ is randomly generated crossover-point		
 		\STATE $Cr_1 = [\Psi_i(1:cp), \Psi_{rn}(cp +1:p)]$
 		\STATE $Cr_2 = [\Psi_{rn}(1:cp), \Psi_{i}(cp +1:p)]$
 		\STATE $Cr= [Cr, \mu(Cr_1), \mu(Cr_2)]$
 		\STATE $VM^{feasible}$=Feasible VM Allocation($Cr$) 
 		\STATE  $[RU,PW] = \eta(VM^{feasible})$
 		\ENDFOR	
 		\STATE $\Psi^g=[\Psi^g,Cr]$
 		\STATE $[\Psi^{g+1}= Pareto-optimal(\Psi^g)]$
 		\ENDFOR
 		
 	\end{algorithmic}
 \end{algorithm} 
 
 \begin{algorithm}[!htbp]
 	\caption{ Pareto\_optimal($\psi_{g}$)}
 	\label{pareto}
 	\begin{algorithmic}[1]
 		\FORALL { $i^{th}$ VM allocation belongs to generation $g$ i.e. $\psi_i^g$}
 		\STATE Initialize $domset_i= \emptyset$, $domcount_i=0$
 		\FORALL {$j^{th}$  VM allocation$\psi_j^g$}
 		\IF {$j^{th}$ VM allocation dominates $i^{th}$ VM allocation i.e. $\psi_i^g \prec \psi_j^g$ }
 		\STATE $domset_i$ =$domset_i \cup \psi_j^g$
 		\ELSIF { $\psi_j^g \prec \psi_i^g$ }
 		\STATE $domcount_i$ =$domcount_i + 1$
 		\ENDIF
 		\IF {$domcount_i == 0$} 
 		\STATE Assign rank to $i^{th}$ VM allocation and initialize first front, as: $Rank[\psi_i^g]$=1, $Front_1=Front_1 \cup \psi_i^g$
 		\ENDIF
 		\ENDFOR
 		\ENDFOR
 		\STATE $current = 1$
 		\WHILE {$Front_{current} \neq \emptyset $}
 		\STATE $nextFront = \emptyset$
 		\FORALL {$i^{th}$ VM allocation in current Front, i.e. $\psi_i^g \in Front_{current}$}
 		\FORALL {$j^{th}$ VM allocation in $i^{th}$ dominant set, $\psi_j^g \in domset_i $}
 		\STATE  Decrement $j^{th}$ dominant count $domcount_j$ =$domcount_j - 1$
 		\IF {$domcount_j==0$}
 		\STATE Assign rank to $j^{th}$ VM allocation and initialize next front, as: $Rank[\psi_j^g]=1 + i, nextFront = nextFront \cup \psi_j^g$
 		\ENDIF
 		\ENDFOR 
 		\ENDFOR
 		\STATE 	 $current = current + 1$, $Front_{current}= nextFront $
 		\ENDWHILE
 		\RETURN {0}	
 	\end{algorithmic}
 \end{algorithm} 
\section{ Resource management algorithm and complexity analysis}  Algorithm \ref{Proposed_algo} is the main module that describes the overall operational summary of proposed energy-efficient resource management approach, which executes periodically to manage the elastic resources without any external intervention. 
\begin{algorithm}[!htbp]
	\caption{Proposed resource management: main algorithm ()}
	\label{Proposed_algo}
	
	\begin{algorithmic}[1]
		\FOR {each $j^{th}$ server $s \in S$}
		\FOR {each $i^{th}$ VM on server $s_j$}		
		\STATE Forecast resource utilization of each task as: $task_i^{pred.CPU}$, $task_i^{pred.Mem}\Leftarrow Algorithm 1() + EDP$  
		\ENDFOR
		\ENDFOR
		\STATE Apply K-Means clustering to group predicted tasks according to their resource utilization
		\STATE Map different clusters that fits to particular VM type by applying Eq. \ref{eq.autoscale}
		\STATE After mapping, determine required number of VMs of particular type = Number of predicted tasks in the selected cluster 
		\STATE CALL Multi-objective VM placement(selected $v$)
		\STATE When actual task arrives, assign them to selected auto-scaled VM
		\STATE Repeat above steps for each prediction interval
		\end{algorithmic}
\end{algorithm} 
The resource requirement of future tasks for each VM on a server is predicted in lines 1-5 by calling Algorithm \ref{algo-prediction}, which forecasts the resource utilization for the next session. Algorithm \ref{algo-prediction} provides training operation steps for OM-FNN prediction system, whose time complexity depends on size of neural network (L), number of networks (N), number of input nodes (n), which becomes $ O(n^2 NL) $. Line 6 calls K-Means clustering with Elbow method, whose time complexity comes out to be $O(K*q)$, where $K$ is number of clusters and $q$ is the number of tasks. Lines 7 and 8 provide steps for VM Autoscaling. Line 9 calls multi-objective VM placement module, provided in Algorithm \ref{algo-mob-lb}
that works on $n$ number of solutions and number of generations ($Gmax$), servers (p), VMs (q). It further calls Pareto-optimal module i.e. Algorithm \ref{pareto}, whose time consumption comes out to be $O(n^2 \times o)$ where $o$ is number of objectives. Hence, overall, time complexity becomes $O(on^2pqKGmax)$.

\section{Performance evaluation}

\subsection{Experimental set-up}
The simulation experiments are executed on a server machine assembled with two Intel\textsuperscript{\textregistered} Xeon\textsuperscript{\textregistered} Silver 4114 CPU with 40 core processor and 2.20GHz clock speed. The computation machine is deployed with 64-bit Ubuntu 18.04 LTS, having main memory of 128 GB. The data center environment was set up with three different types of server and four types of VMs configuration shown in Tables \ref{table:server} and \ref{table:vm} in Python version-3. The resource features like power consumption ($P_{max}, P_{min}$), MIPS, RAM and memory are taken from real server IBM \cite{IBM1999} and Dell \cite{Dell1999} configuration where $S_1$ is 'ProLiantM110G5XEON3075', $S_2$ is 'IBMX3250Xeonx3480' and $S_3$ is 'IBM3550Xeonx5675'. Furthermore, the experimental VM configuration are inspired from the VM instances from Amazon website \cite{amazon1999EC2}. The description of different parameters and their values used for OM-FNN predictor training are listed in Table \ref{table:name1}
\begin{table}[!htbp]
	\centering
	
	\caption[Table caption text] {Server Configuration}  
	\label{table:server}
	\resizebox{9cm}{!}{
		\begin{tabular}{lccccccc}
			\hline
			Server&PE&MIPS&RAM(GB)&Memory(GB)&$PW_{max}$&$PW_{min}$/$PW_{idle}$\\
			\hline
			$S_1$ 	& 2&2660&4&160&135&93.7 \\
			$S_2$	& 4&3067&8&250&113&42.3 \\
			$S_3$	& 12&3067&16&500&222&58.4 \\
			
			
			\hline
	\end{tabular}}
\end{table}

\begin{table}[!htbp]
	\centering
	
	\caption[Table caption text] {VM configuration}  
	\label{table:vm}
	\begin{tabular}{lcccc}
		\hline
		VM type& PE &MIPS&RAM(GB)&Memory(GB)\\
		\hline
		$v_{small}$ ($v_S$)&1&500&0.5&40\\
		$v_{medium}$ ($v_M$)&2&1000&1&60\\
		$v_{large}$ ($v_L$)&3&1500&2&80\\
		$v_{Xlarge}$ ($v_{XL}$)&4&2000&3&100\\

		\hline
	\end{tabular}
\end{table}

\begin{table}[!htbp]
	\centering
	
	\caption[Table caption text] {Experimental set-up parameters for training of OM-FNN and their values.}  
	\label{table:name1}
	\begin{tabular}{lr}
		\hline
		Parameter    & Value  \\
		\hline
		Number of nodes in each set & two nodes: CPU and memory\\
		Input neural set of nodes ($n$)     & 3    \\
		Hidden layer set of nodes ($p$)        & 5        \\
		Output layer set of nodes ($q$)      & 1          \\
		Maximum epochs ($G_{max}$)     & 200        \\
		Size of training data & 80\%            \\
		Number of population  & 10\\
		\hline
	\end{tabular}
\end{table}
The resource utilization for different VMs follow the traces from publicly available real workloads including Google Cluster Data (GCD) dataset. GCD has resources CPU, memory, disk I/O request and usage information of 672,300 jobs comprised of one or more tasks executed on 12,500 servers for the period of 29 days \cite{reiss2011google}. The resource utilization information has been used in two forms. In the first form, CPU and Memory utilization of VMs are consolidated over different time-intervals, including 5 min, 10 min, ..., 60 min, 1440 min. In the second form, the percentage of CPU and memory utilization of a VM is aggregated in every five minutes over a period of twenty-four hours, which is taken as resource requirement of a task. Following performance metrics are evaluated: (i)Accuracy of Predicted Workload vs Actual Workload, (ii) comparison of Single Input and Single Output Resource Neural Network (SISO-FNN) and OM-FNN based prediction, (iii) comparison of proposed near optimal and optimal VM autoscaling, (iv) resource utilization, power consumption and number of active servers obtained by multi-objective VM placement (v) overall improvement in resource utilization and power saving achieved by proposed approaches.
	
	

\subsection{Comparative data conceptual analysis: }The proposed work is compared according to different performance metrics with various state-of-art approaches including SaDE \cite{kumar2018workload} and Backpropagation \cite{prevost2011prediction} for resource prediction.
\begin{itemize}
\item \textit{Self-adaptive DE (SaDE) } \cite{kumar2018workload}: SaDE works on multiple solutions and explores optimal solution by searching in multiple directions. However, the adaptation feature is restricted to mutation selection only in case of SaDE. 
\item \textit{Backpropagation (BP) } \cite{prevost2011prediction}: BP is a commonly used learning algorithm for neural networks that works on single solution and provides supervised learning by minimizing the difference between actual and predicted output with respect to gradient descent over the iterations.    
\end{itemize}
 The power consumption and resource utilization are compared with Genetic Algorithm based secure VM placement \cite{singh2019secure}, Random-Fit \cite{jung2010mistral} and Best-Fit heuristic \cite{beloglazov2012optimal}.
\begin{itemize}
	\item \textit{Secure and Energy efficient load balancing (SEA-LB) } \cite{singh2019secure}: The VMs are placed with three different objectives including maximum resource utilization and minimum; power consumption and side-channel attacks in SEA-LB by applying modified genetic algorithm approach. The security is provided by minimizing the number of shared servers at the cost of resource utilization. 
	\item \textit{Random-Fit (RF)} \cite{jung2010mistral}: The VMs are placed on  randomly selected servers that satisfies their resource requirement. This strategy works on randomly generated single solution, so there is no further scope of optimization. It can easily suffer from performance degradation due to rise of over/under-loaded servers. RF-VMP is implemented with fixed size VMs as mentioned in Table \ref{table:vm} without prediction and autoscaling for comparison. 
	\item \textit{Best-Fit (BF)} \cite{beloglazov2012optimal}: Similarly, Bf-VMP is implemented (without resource prediction and VM autoscaling) such that fixed size VMs are placed on best servers with minimum resource capacity that can satisfy the VM's resource requirement to avoid resource wastage. This VMP also works on single solution and selection of best server for each VM is a time consuming process.   
\end{itemize}
 
\subsection{ Multiple resource prediction}
\begin{figure*}[!htbp]
	\centering

   	\subfloat[GCD-CPU on SISO-FNN (PWS=10min)\label{subfig-1:dummy}]{%
   	\includegraphics[width=0.47\textwidth]{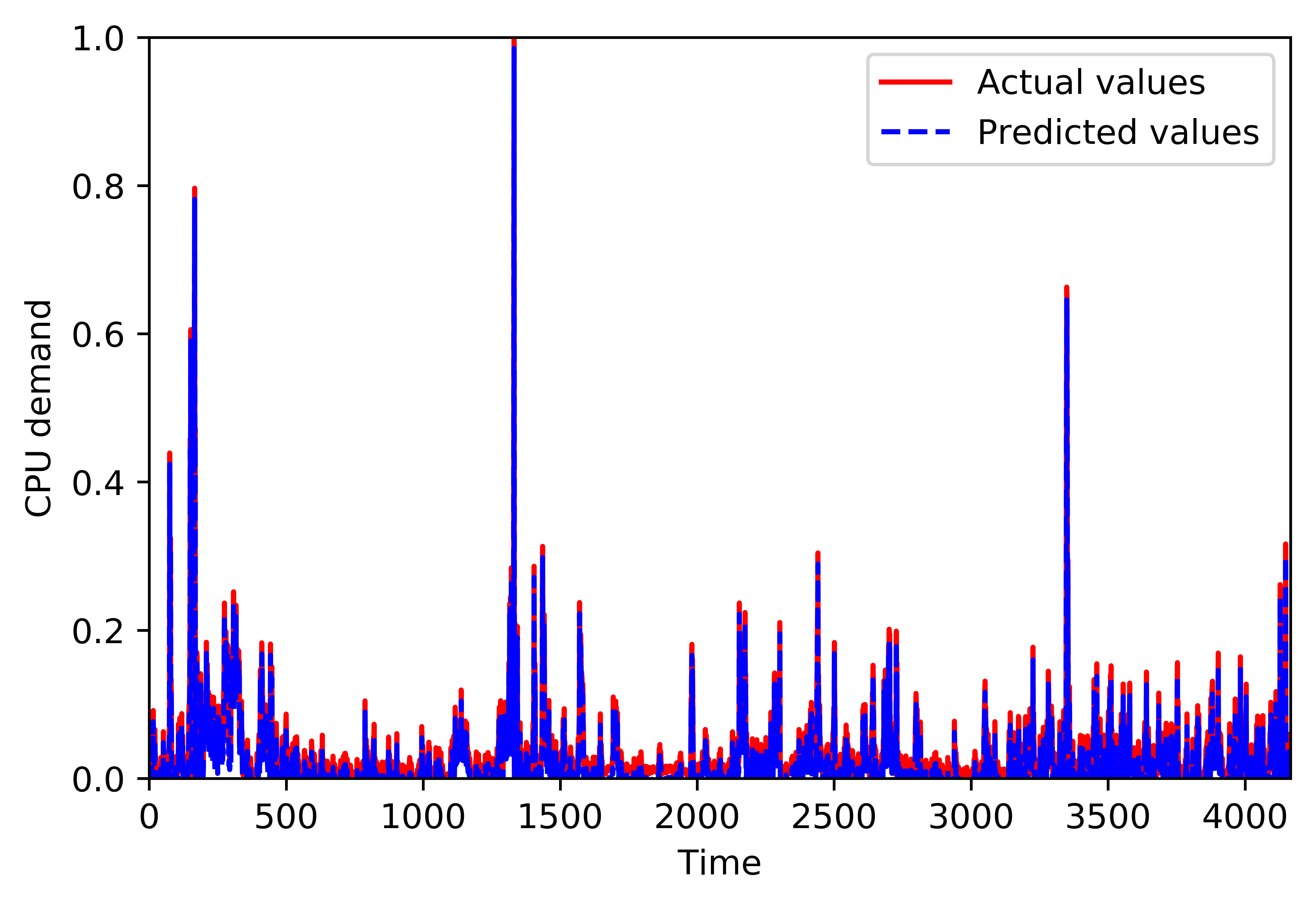}
   }
	\hfill
\subfloat[GCD-CPU on SISO-FNN (PWS=60min)\label{subfig-1:dummy}]{%
	\includegraphics[width=0.47\textwidth]{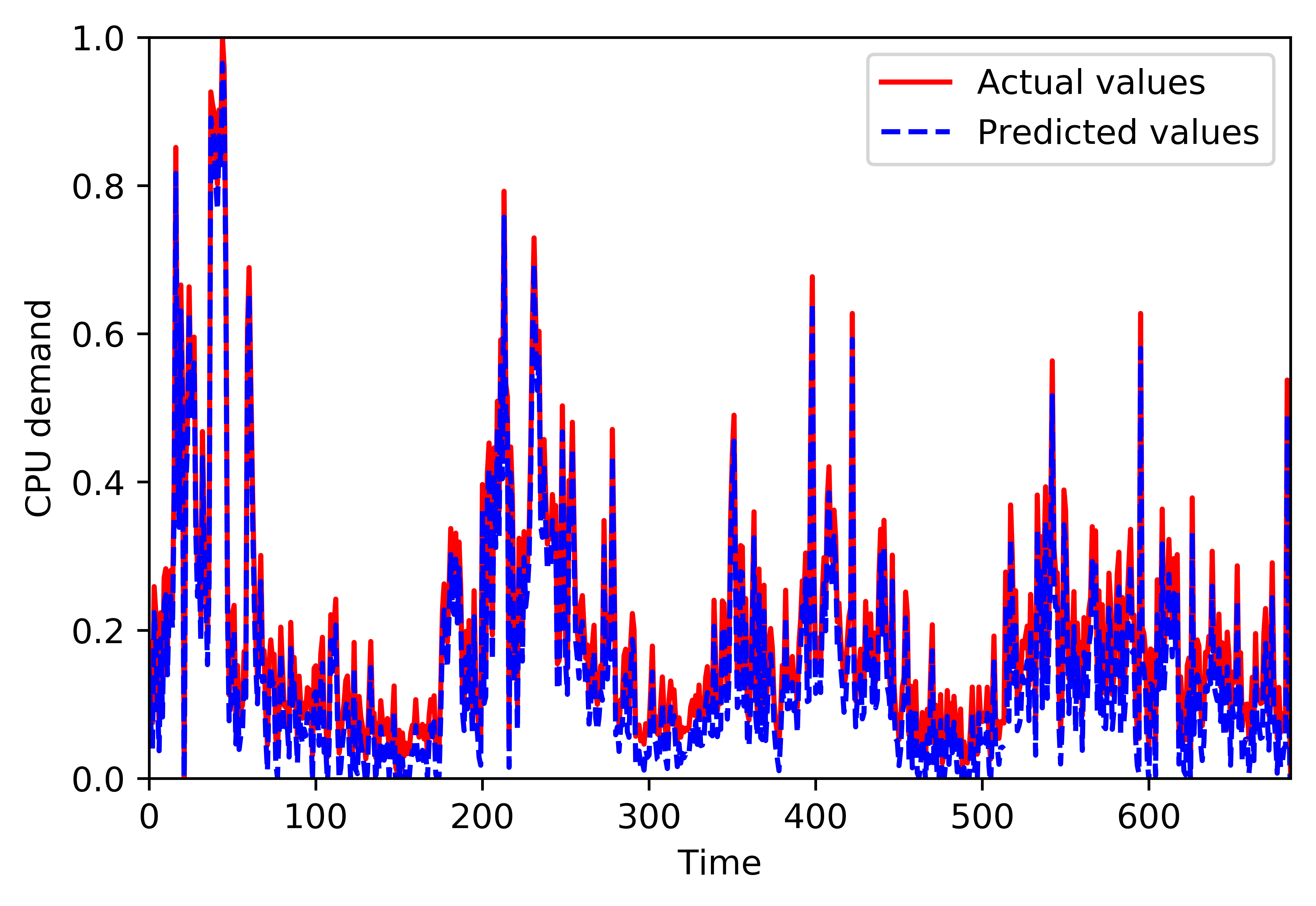}
}

	\hfill
	\subfloat[GCD-memory on SISO-FNN (PWS=10 min)\label{subfig-2:dummy}]{%
		\includegraphics[width=0.47\textwidth]{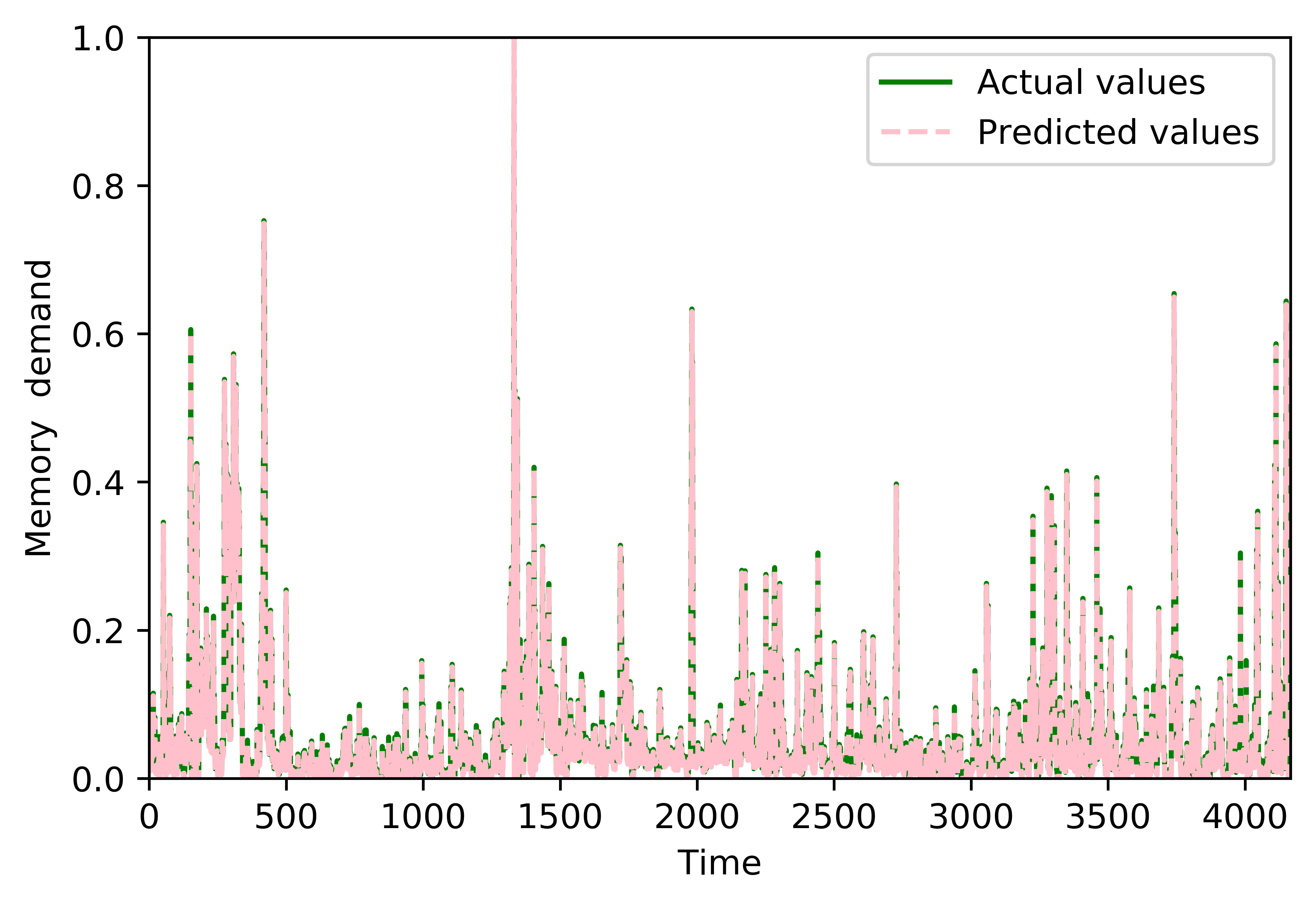}
	}
\hfill
\subfloat[GCD-memory on SISO-FNN (PWS=60 min)\label{subfig-2:dummy}]{%
	\includegraphics[width=0.47\textwidth]{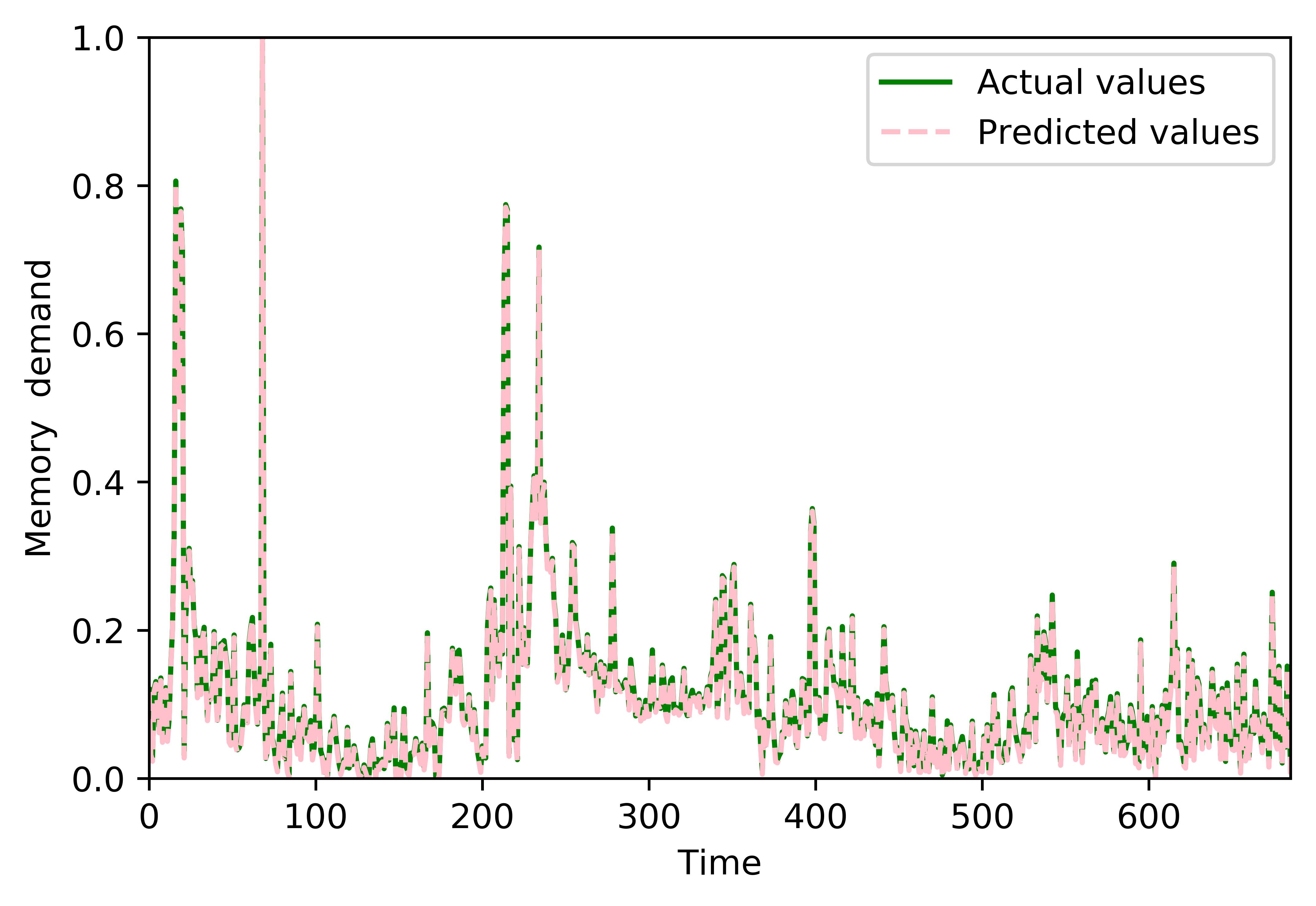}
}
   \hfill
  	\subfloat[GCD-CPU \& GCD-memory on OM-FNN (PWS=10 min)\label{omrnn-10min}]{%
  	\includegraphics[width=0.47\textwidth]{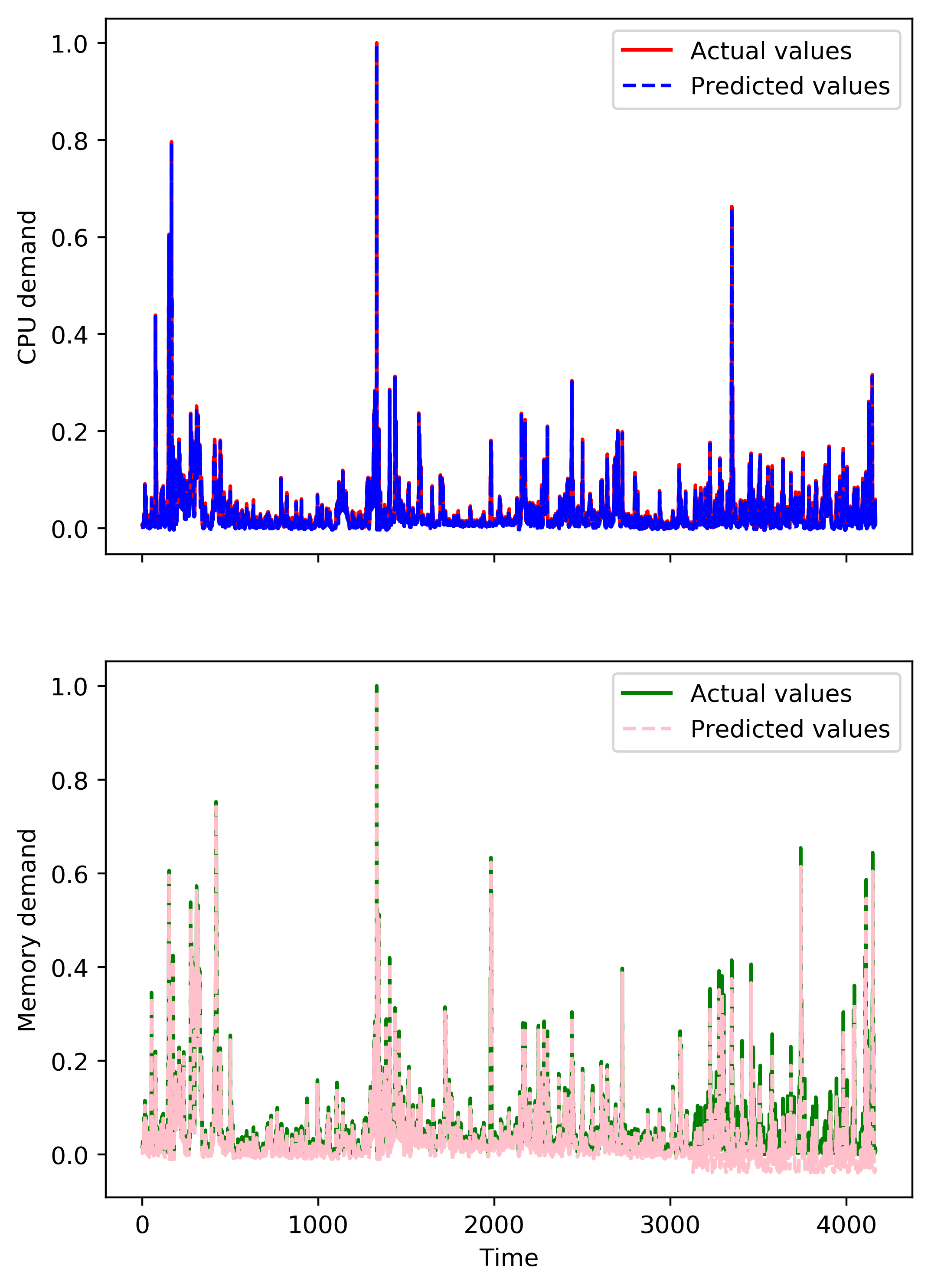}
  }
	\hfill
	\subfloat[GCD-CPU \& GCD-memory on OM-FNN (PWS=60 min)\label{omrnn-60min}]{%
	\includegraphics[width=0.47\textwidth]{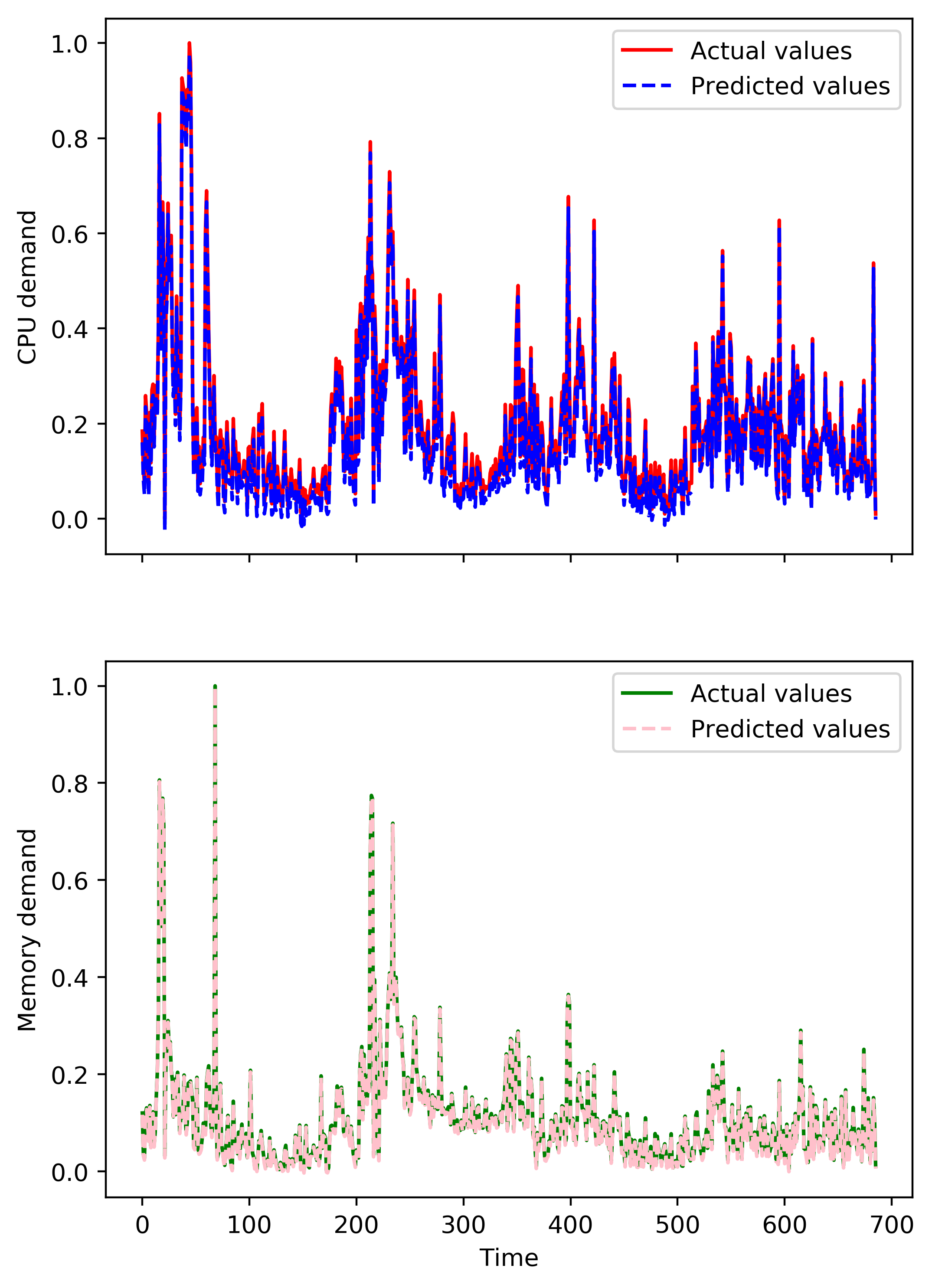}
}

	\caption{Comparison of predicted vs actual CPU and memory utilization for SISO-FNN and proposed OM-FNN based prediction system}
	\label{omrnn}
	
\end{figure*}

The performance evaluation of the proposed framework initiates with the investigation of accuracy of SISO-FNN and OM-FNN prediction system, both are optimized with proposed TaDE algorithm. Its effectiveness can be seen in Fig. \ref{omrnn} (where PWS is 'Prediction Window Size' means prediction interval) that predicted CPU and memory usage have almost overlapped with actual resource usage for GCD workloads. Fig. \ref{omrnn} shows comparison of predicted versus actual CPU and memory utilization for SISO-FNN and OM-FNN based predictor system over prediction interval of 10 and 60 minutes.

 Table \ref{table:siso-omrnn} compares SISO-FNN and OM-FNN subject to accuracy, operational time and space consumption. The accuracy of OM-FNN prediction system is measured in terms of prediction error obtained by applying Eq. \ref{rmse}, which is lesser or equal to prediction error of SISO-FNN for almost every case. However, the operational time (in msec) and space (in Bytes) consumption in case of OM-FNN is always lesser than that of SISO-FNN. The time and space requirement have been reduced upto 46.8\% and approx. 8.4\% respectively in case of OM-FNN as compared to SISO-FNN. This is because in case of SISO-FNN, $N$ networks of size $L$ are separately initialized for each resource (number of resources are two for reported results) which consumes excess memory space and time, while in case of OM-FNN, these networks are initialized only once for any number of resources that reduce time and space consumption overhead. In addition, it is observed that with increase in size of PWS or prediction interval, the execution time and space requirement decreases and prediction error increases. The reason is that with bigger PWS, the prediction model under training get lesser number of data samples for pattern recognition or learning than smaller prediction interval for the same benchmark dataset. 
\begin{table}[!htbp]
	
	\caption[Table caption text] {OM-FNN vs SISO-FNN Predictor: Comparison of CPU and Memory RMSE, Time and Memory elapsed . }  
	\label{table:siso-omrnn}
	\small
	\resizebox{12cm}{!}{
	\centering
	\begin{tabular}{ |l|llll|llll| }
		
			\hline
			PWS&\multicolumn{4}{c}{SISO-FNN Predictor}&\multicolumn{4}{|c|}{OM-FNN Predictor} \\
			\cline{2-9}
			(min) &$\xi_{CPU}$&$\xi_{Mem}$&Time (ms)&Memory (B)&$\xi_{CPU}$ &$\xi_{Mem}$&Time (ms)&Memory (B)\\
			\hline
			10&0.0022&0.0044&176.47&9.43E+07&0.0011&0.0036&118.78&9.12E+07\\
			20&0.0039&0.0094&111.81&1.01E+08&0.0020&0.0062&111.64&9.00E+07\\
			30&0.0116&0.0981&100.92&9.24E+07&0.0135&0.0106&90.78&9.06E+07\\
			60&0.0225&0.022&22.46&8.93E+06&0.0022&0.0038&25.08&8.99E+06\\
			1440&0.033&0.046&1.28&8.86E+06&0.0311&0.0230&0.643&8.12E+06\\

			\hline
		\end{tabular}}
		
\end{table}

\par Table \ref{table:comparison_prediction} shows prediction error comparison of proposed TaDE algorithm trained OM-FNN prediction model, Backpropagation (BP) \cite{prevost2011prediction} and Self-adaptive Differential Evolution (SaDE) (\cite{kumar2018workload}) algorithm trained SISO-FNN prediction models. The comparison verifies that the proposed approach reduces the forecast error up to 95.18\% and 85.29\% against BP and SaDE approach respectively for GCD-CPU utilization logs for prediction interval of 5 minutes. Similarly, the forecast error of  GCD-Memory utilization is reduced up to 96.25\% and 78.04\%, against BP and SaDE based approaches respectively. The reason behind an improved accuracy is the three phase adaptation at mutation, crossover and control parameter tuning phases. It allows the selection of most appropriate mutation operator and crossover operator for a particular population vector according to $msp$ and $csp$ probability values along with adaptation of learning period to update crossover and mutation rate. These operators help in reaching closer to the global optimal solution while learning process of OM-FNN predictor.

\begin{table}[!htbp]
	\centering
	\caption{Prediction error (RMSE) comparison of Back Propagation, SaDE and TaDE for Google Cluster Dataset}
	\label{table:comparison_prediction}
		\resizebox{9cm}{!}{
	\begin{tabular}{@{\extracolsep{4pt}}ccccccc@{}}  
		\toprule
		\multirowcell{2}{Prediction \\ Time (min)}	& \multicolumn{2}{c}{BP\cite{prevost2011prediction} } & \multicolumn{2}{c}{SaDE \cite{kumar2018workload}}  &\multicolumn{2}{c}{Proposed TaDE }\\ 
		\cline{2-3} \cline{4-5} \cline{6-7} 
		&	CPU		&	Memory		&	CPU		&	Memory		&	CPU 		&  Memory \\ \midrule
		5 	& 0.0087	& 0.019	& 0.0017	& 0.0023 	& 0.00025	& 0.0010\\
		10 	& 0.014	& 0.024 	& 0.0025	& 0.0041	& 0.0021	& 0.0009\\
		20 	& 0.026		& 0.048	& 0.0041	& 0.0055	& 0.0028	& 0.0033\\
		30  & 0.043		& 0.051	& 0.0055	& 0.0095	& 0.0078	& 0.0096\\
		60  & 0.103		& 0.087	& 0.010		& 0.012 	& 0.0099	& 0.0198\\
		\lasthline
	\end{tabular}}
\end{table}
\par
\begin{figure*}[!htbp]
	\centering

	\subfloat[Learning period adaptation vs Number of epochs \label{fig:crossovermutationlearningrate}]{%
		\includegraphics[width=0.48\textwidth]{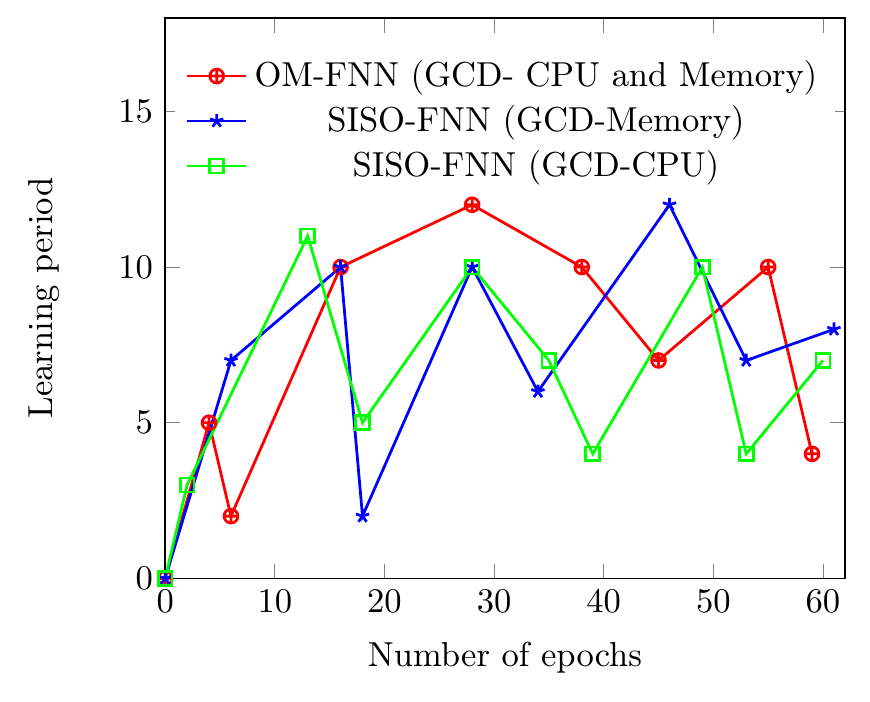}
	}
		\hfill
	\subfloat[Comparison of effect of Tri-phase adaptation and SaDE on convergence  \label{fig:convergencecomparison}]{%
		\includegraphics[width=0.48\textwidth]{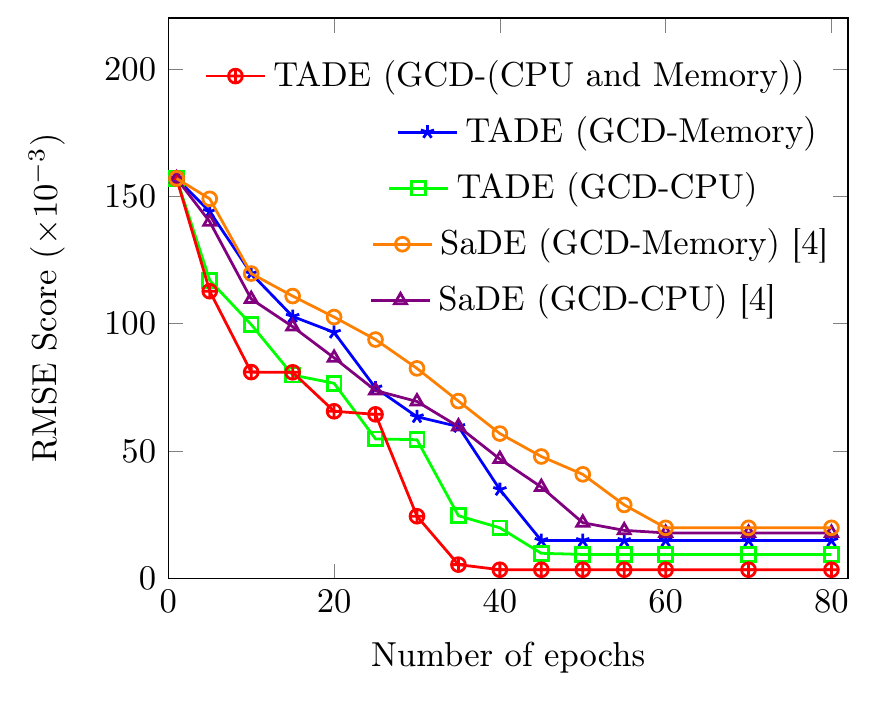}
	}	
	\caption{Comparison of SISO-FNN and OM-FNN based prediction}
	\label{tade}
	
\end{figure*}

 
Fig. \ref{fig:convergencecomparison} shows the effect of tri-phase adaptation on convergence rate of GCD-CPU, GCD-Memory on SISO-FNN and GCD-(CPU and Memory) on OM-FNN for PWS of 10 minutes. The learning of OM-FNN predictor has become faster due to three stage optimization adaptation, which is lesser than 45 iterations for both prediction systems. It shows that TaDE based approach converges faster and produce better accuracy, as compared to single phase adaptation i.e. SaDE based prediction approach. This is due to the fact that TaDE incorporates an additional control parameter tuning adaptation, which guides optimized learning and updates crossover and mutation rate only when there appears no significant improvement during evolution. On the other hand, the learning period of mutation rate is fixed in case of SaDE approach \cite{kumar2018workload}, which leads to slower and pre-mature convergence.
The effect of control parameter tuning adaptation is shown in Fig. \ref{fig:crossovermutationlearningrate}, where, learning period of the mutation and crossover rate updates adaptively according to the progress of optimization. This plot shows learning period adaptation obtained during execution of GCD-CPU and Memory utilization log trace of PWS equals to 60 minutes. 

\subsection{Proactive VM Autoscaling}
 The effect of VM autoscaling is analyzed in Table \ref{table:autoscaling}, which shows the selected number and type of VMs scaled during experimental execution, for different number of tasks. In addition, this Table compares the proposed VM autoscaling approach with actual resource utilization which gives optimal scaling of VMs. It is to be noticed that consecutive OM-FNN prediction and VM-autoscaling approach correctly determines required VM type with its (closer) quantity for execution of future tasks in anticipation. The bold numbers highlight the most closer quantity of selected VM type obtained with OM-FNN approach. Therefore, the proposed approach generates closer to optimal solution for scaling of VMs and provides near-optimal VM autoscaling for cloud data center. Otherwise, SLA violations or resource wastage may occur due to an inappropriate scaling of VMs.
\begin{table}[!htbp]
	
	\caption[Table caption text] {Comparison of VM Autoscaling with actual and predicted resource utilization  }  
	\label{table:autoscaling}
	\small
	\resizebox{12cm}{!}{
		\centering
		\begin{tabular}{ |l|l|l| }
			
			\hline
				 Tasks &\multicolumn{2}{c|}{VM Type(No. of VMs)} \\
		     \cline{2-3}
			 &  VM Autoscaling & VM Autoscaling \\
			 & (with actual resource utilization) &(with predicted resource utilization)\\
			\hline
			100&$v_{S}$(83), 	$v_{M}$(11), $v_{L}$(6), $v_{XL}$(0)&$v_{S}$(67), $v_{M}$(17), $v_{L}$(14), $v_{XL}$(2)\\
			200&$v_{S}$(169), 	$v_{M}$(17), $v_{L}$(11), $v_{XL}$(3)&$v_{S}$(143), $v_{M}$(29), $v_{L}$(21), $v_{XL}$(7)\\
			400&$v_{S}$(343), $v_{M}$(31), $v_{L}$(19), $v_{XL}$(7)&$v_{S}$(321), $v_{M}$(\textbf{37}), $v_{L}$(29), $v_{XL}$(13)\\
			600&$v_{S}$(512), 	$v_{M}$(47), $v_{L}$(23), $v_{XL}$(9)&$v_{S}$(\textbf{497}), $v_{M}$(\textbf{49}), $v_{L}$(35), $v_{XL}$(19)\\
			800&$v_{S}$(691), 	$v_{M}$(63), $v_{L}$(31), $v_{XL}$(15)&$v_{S}$(671), $v_{M}$(68), $v_{L}$(\textbf{37}), $v_{XL}$(24)\\
			1000&$v_{S}$(862), $v_{M}$(82), $v_{L}$(39), $v_{XL}$(17)&$v_{S}$(829), $v_{M}$(102), $v_{L}$(\textbf{44}), $v_{XL}$(25)\\
			1200&$v_{S}$(1039), $v_{M}$(97), $v_{L}$(41), $v_{XL}$(23)&$v_{S}$(998), $v_{M}$(112), $v_{L}$(59), $v_{XL}$(31) \\

			\hline
	\end{tabular}}
	
\end{table}

\subsection{VM Placement}
 \par The autoscaled VMs are placed on energy- efficient servers by applying multi-objective VM Placement (VMP). To investigate the efficiency and utility of various algorithms (i.e. OM-FNN based prediction and clustering based VM autoscaling) in proposed approach, four combination of VMP are developed and analyzed. Table \ref{table:VMP} shows experimental results of different combination of VMPs including, online Prediction and Autoscaling based VM Placement (PA-VMP), Optimal Autoscaling based VM Placement (OA-VMP), Prediction Without Autoscaling based VM Placement (PWA-VMP), and Without Prediction and Without Autoscaling based VM Placement (WPWA-VMP) \cite{singh2019secure}. OA-VMP is an exact algorithm where, VM autoscaling is done with exact resource requirement (actual CPU and Memory utilization taken from GCD) of tasks to generate feasible and optimal autoscaling based placement of VMs. PA-VMP firstly, predicts future utilization of multiple resources of each task by applying OM-FNN based prediction approach, secondly, determines number and types of VMs to be scaled and finally, place autoscaled VMs on energy-efficient servers.  
\begin{table}[!htbp]
	
	\caption[Table caption text] {Experimental results of different objectives for four proposed VM placement approaches  }  
	\label{table:VMP}
	\small
	\resizebox{12cm}{!}{
		\centering
\begin{tabular}{ |l|l|cccccc| }
	\hline
	& & \multicolumn{6}{ c| }{Size of Data center (Number of VMs)} \\
	\cline{3-8}
	Approach & Objectives & 200 & 400&600&800&1000&1200 \\ \hline
	
	\multirow{3}{*}{OA-VMP} & RU (\%) &71.56&69.45&70.58&71.69&72.7 &71.06   \\
	& PW (W)& 2.75E+3& 5.11E+3& 7.53E+3& 10.22E+3& 12.70E+3& 15.20E+3\\
	& APMs &22&41&60&79&100& 119\\

	\hline
	\multirow{3}{*}{PA-VMP} & RU (\%) &67.70&66.89&67.75&66.51&67.95 &68.94   \\
	& PW (W)& 3.18E+3& 7.30E+3& 10.34E+3& 13.15E+3& 16.95E+3& 20.26E+3\\
	& APMs &34&77&97&112&168& 181\\
	\hline
	\multirow{3}{*}{PWA-VMP} & RU (\%) &64.98&64.57&64.05&63.34&63.87 &63.32   \\
	& PW (W)& 8.80E+3& 19.6E+3& 21.7E+3& 33.1E+3& 35.9E+3& 39.2E+3\\
	& APMs &69&109&137&199&269& 334\\
	
	\hline
	\multirow{3}{*}{WPWA-VMP} & RU (\%) &59.08&58.57&59.05&58.34&58.87 &59.32   \\
	& PW (W)& 23.99E+3& 37.9E+3& 48.14E+3& 51.25E+3& 56.50E+3& 58.12E+3\\
	& APMs &110&189&287&397&509& 654\\

	\hline
\end{tabular}}
\end{table}
 
To analyze sole efficiency of online prediction, we develop PWA-VMP, which, is the combination of consecutive online prediction of resource utilization and energy-efficient VM placement, where VM autoscaling step is skipped. WPWA-VMP is developed, in order to test the isolated performance of VMP. 

\par The comparison given in Table \ref{table:VMP} shows performance increases in the trend: $OA-VMP \ge PA-VMP \ge PWA-VMP \ge WPWA$ subject to maximum resource utilization, minimum power consumption and number of active servers. For each combination of VMP, very slight or no variation is observed in resource utilization with increasing size of data center. On the other hand, both power consumption and number of active servers grows with rise in size of data center. The proposed multi-objective VMP is able to achieve resource utilization upto 59.32\% with possible reduction in power consumption and number of active servers. The application of online prediction before VMP (i.e. PWA-VMP), has enhanced its performance with improvement in resource utilization up to 5.66\%, reduction in power consumption and number of active server up to 33.9\% and 48.9\% respectively. The performance of PWA-VMP is further improved by incorporating clustering based VM autoscaling i.e. OA-VMP, which has improved resource utilization up to 7.1\%, minimized power consumption and number of active server up to 60.9\% and 64.3\% respectively. However, as compared to optimal VMP (i.e. PA-VMP), OA-VMP allows some lesser performance, in which, the resource utilization (\%) is lesser by 3.76\%, reduction in power consumption and number of active servers is lesser by 1.33 times and 1.52 times respectively, of that of OA-VMP. Hence, the PA-VMP allows near-optimal performance, very much closer to the optimal solution.

\begin{figure}[!htbp]
	\centering
	\includegraphics[width=0.7\linewidth]{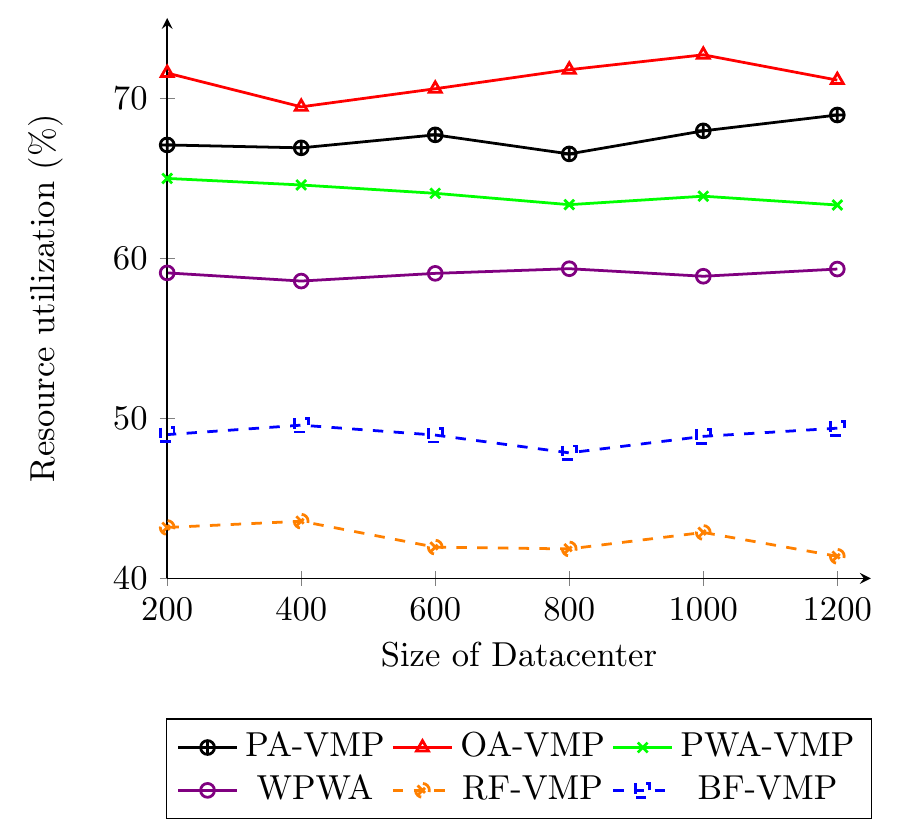}
	\caption{Comparison of Resource utilization}
	\label{fig:ru}
\end{figure}

\begin{figure}[!htbp]
	\centering
	\includegraphics[width=0.7\linewidth]{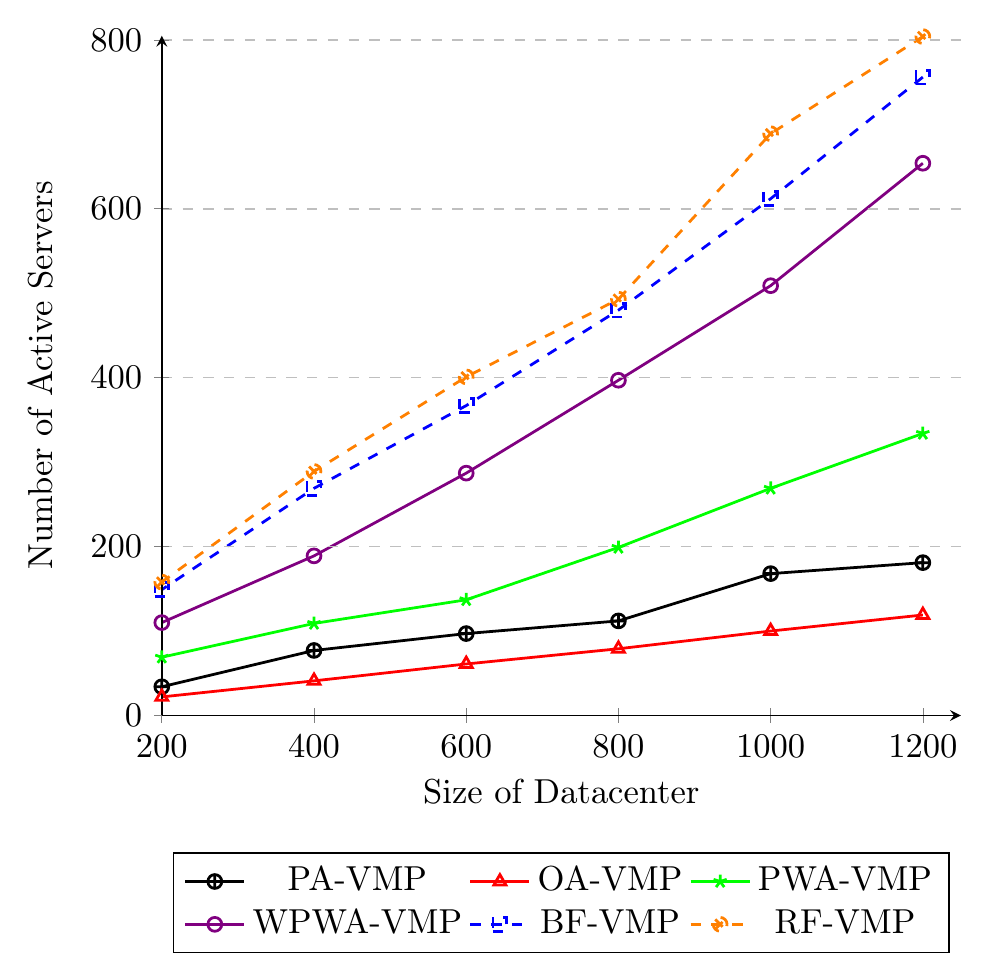}
	\caption{Comparison of number of Active PMs}
	\label{fig:apms}
\end{figure}

\begin{figure}[!htbp]
	\centering
	\includegraphics[width=0.7\linewidth]{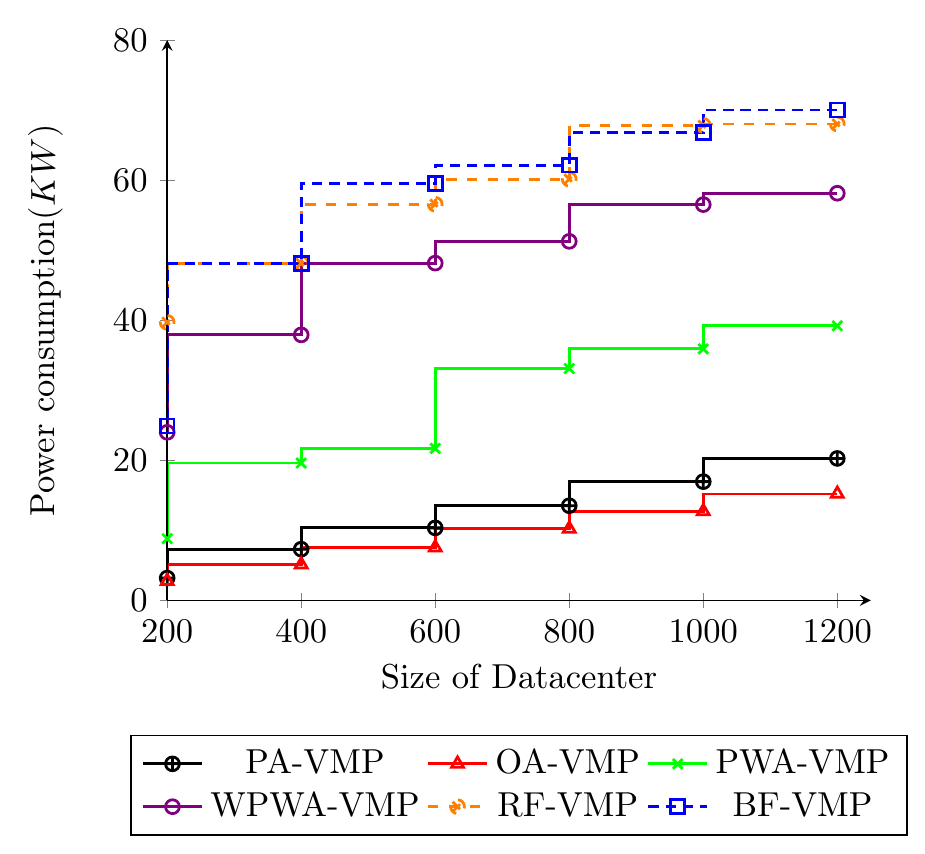}
	\caption{Comparison of Power consumption}
	\label{fig:power}
\end{figure}

Fig. \ref{fig:ru} compares resource utilization of PA-VMP and shows improvement up to 60.97\%, 38.7\% against existing RF-VMP \cite{jung2010mistral} and BF-VMP \cite{beloglazov2012optimal} approaches. Similarly, improvement in resource utilization is achieved up to 73.6\%, 48.1\% for OA-VMP, up to 54.8\%, 32.3\% for PWA-VMP and up to 42.7\%, 21.8\% for WAWP-VMP \cite{singh2019secure} (which is genetic algorithm based secure VMP) respectively, over RF-VMP, BF-VMP approaches. In addition, the number of active servers are compared in Fig. \ref{fig:apms}, which increases with increase in the size of data center. It is observed that least number of servers are active in case of optimal autoscaling followed by energy efficient VM placement approach and random-fit VM placement requires highest number of active servers for the same size of data center. The number of active servers in case of OA-VMP, PA-VMP, PWA-VMP and WAWP-VMP, are scale down by 84.3\%, 76.1\%, 55.8\%, and 13.5\% respectively, against BF-VMP. The reduction in number of active servers by applying OA-VMP, PA-VMP, PWA-VMP and WAWP-VMP are 85.1\%, 77.4\%, 58.5\% and 18.6\% respectively, over RF-VMP approach. The comparison of power consumption is shown in Fig. \ref{fig:power} where proposed energy-efficient PA-VMP approach scales down power consumption by (70.7-89.8)\% and (70.2-91.1)\% over Best-fit and Random-fit heuristic based VM allocations. The power consumption has raised with increasing size of data center (VMs) and number of active PMs. Furthermore, it is noted that proposed PA-VMP provides near optimal performance for each objective, as compared to optimal VM placement solution using exact algorithm (OA-VMP).

The reason for such improved resource utilization is that proposed MLB applies evolutionary optimization approach that works on $N$ number of solutions and searches for the most optimal and feasible VM allocation, among multiple VM allocations. On the other hand, RF and BF heuristics are bin-packing algorithms which finds out single solution that fit according to the concept of heuristic. Moreover, it is difficult to attain pareto-optimal solution that can satisfy (non-dominated) multiple constraints simultaneously with these heuristics.

\begin{figure*}[!htbp]
	\centering
	\subfloat[Power saving  \label{fig:powersaving}]{%
		\includegraphics[width=0.47\textwidth]{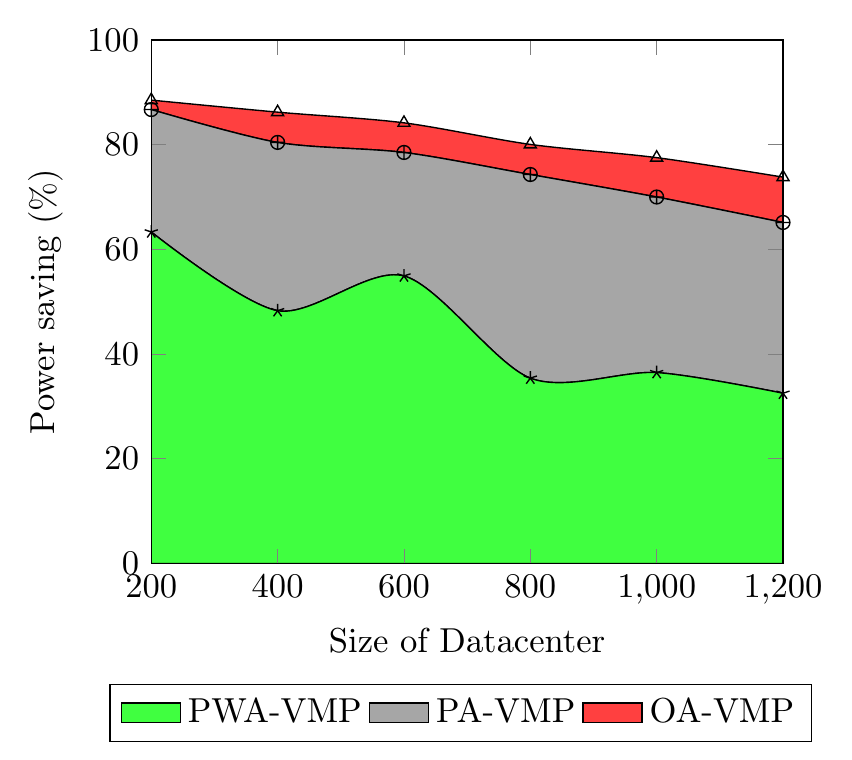}
	}
	\hfill
	\subfloat[Improvement in resource utilization \label{fig:RUImprovement}]{%
		\includegraphics[width=0.47\textwidth]{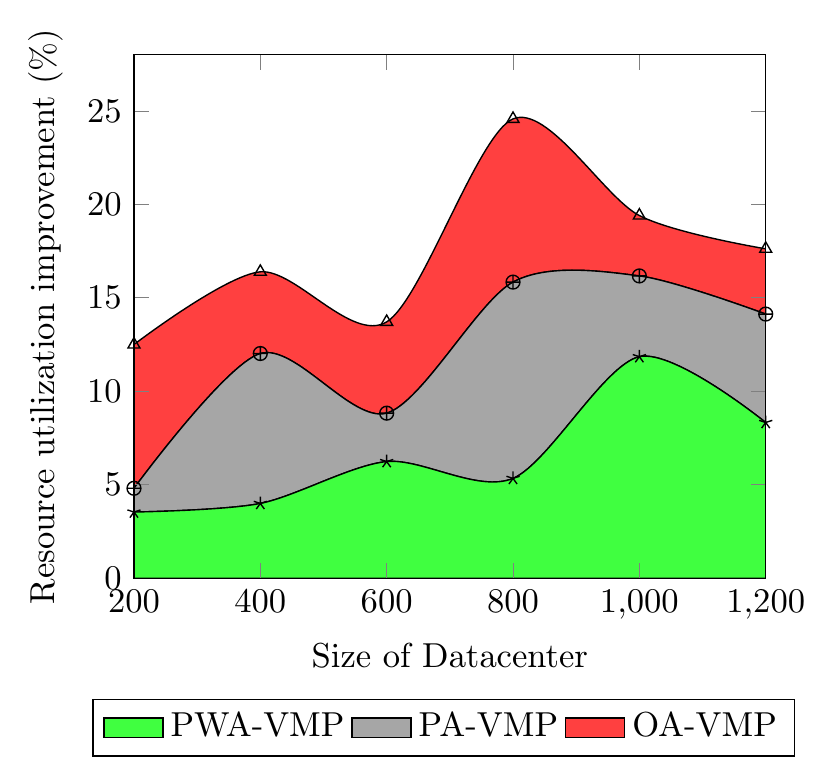}
	}
	\caption{Comparison of performance improvement achieved by proposed resource provisioning approaches over without task prediction and VM autoscaling approach}
	\label{saving}
	
\end{figure*}

Fig. \ref{saving} compares overall improvement in performance per data center for GCD workload, which is achieved by applying proposed approaches including PA-VMP, OA-VMP and PWA-VMP over WPWA-VMP, where VMP is energy-efficient VM placement (described in Algorithm \ref{algo-mob-lb}). Fig \ref{fig:powersaving} shows that power saving has been improved upto 63.33 \%, 86.71\% and 88.5\% by applying PWA-VMP, PA-VMP and OA-VMP respectively, against WPWA-VMP approach. Likewise, Fig \ref{fig:RUImprovement} shows that PWA-VMP, PA-VMP and OA-VMP have improved resource utilization up to 9.9\%, 13.52\% and 21.12\% respectively against WPWA-VMP approach. Furthermore, it is observed that power saving percentage is highest for data center with 200 VMs and then slowly decreases with the increase in the size of the data center. This is due to the accurate prediction of resource requirement of future tasks, followed by efficient autoscaling of VMs and placement of VMs on energy-efficient servers. As a consequence, there is substantial reduction in the number of active servers and VM migrations which have enabled great power saving. 

\section{Conclusion}
In order to reduce the operational cost, resource wastage and conserve power, cloud service providers must strive to use the available physical resources efficiently. From the perspective of cloud service provider, an integrated proactive autoscaling and allocation of VMs approach is proposed, that allows consolidation of load on as few physical machines as possible, without affecting user application performance. A novel OM-FNN predictor is developed to forecast utilization of multiple resources simultaneously with lesser time and space complexity as compared to multiple conventional neural networks. The future applications are grouped into clusters based on their predicted resource requirement and adequate number and type of VMs suitable to a cluster are scaled automatically. The selected autoscaled VMs are placed on energy-efficient servers by applying proposed multi-objective VM placement algorithm. The consecutive resource prediction, autoscaling and placement of VMs, successively reduces resource wastage and saves energy, allowing efficient resource provisioning and management at cloud data center. The proposed integrated approach is thoroughly tested with real resource utilization traces of Google Cluster Dataset. The observed results reveal its superiority over existing methods in terms of prediction accuracy, resource utilization and reduction in power consumption. Despite of above mentioned benefits of the proposed resource management framework, it suffers from limitation of manual selection of number of sets of nodes in the input and output layer of OM-FNN predictor. In future, the predicted tasks can be scheduled on closer placed VMs according to their inter-dependency in order to reduce network traffic and energy expenditure due to communication-intensive VMs while considering resource management decisions.


\bibliography{mybibfile}
\end{document}